\documentclass[%
 reprint,
 amsmath,amssymb,
 aps,
prd,
]{revtex4-2}

\usepackage{graphicx}
\usepackage{dcolumn}
\usepackage{bm}
\usepackage[mathlines]{lineno}
\usepackage{hyperref}
\usepackage{cleveref}[capitalize]

\newcommand{\mtot}{m_{1}+m_{2}}
\newcommand{\msun}{\mathrm{M}_{\odot}}
\newcommand{\dL}{d_{L}}
\newcommand{\qstat}{\hat{q}}
\newcommand{\qstath}{\hat{q}_{\mathrm{H1}}}
\newcommand{\qstatl}{\hat{q}_{\mathrm{L1}}}
\newcommand{\ntrain}{N_{\mathrm{train}}}
\newcommand{\nval}{N_{\mathrm{val}}}

\newcommand{\deepsnr}{\textsc{DeepSNR}}

\mathchardef\mhyph="2D

\begin{document}

\title{
\textsc{DeepSNR}: A deep learning foundation for offline gravitational wave detection
}

\author{Michael Andrews$^{1}$}
\email{michael.andrews@cern.ch}
\author{Manfred Paulini$^{1}$}
\affiliation{$^{1}$ Carnegie Mellon University, Department of Physics, Pittsburgh, Pennsylvania 15213, USA}

\author{Luke Sellers$^{2}$}
\author{Alexey Bobrick$^{2,3,4}$}
\author{Gianni Martire$^{2}$}
\author{Haydn Vestal$^{2}$}
\affiliation{$^{2}$
 Advanced Propulsion Laboratory at Applied Physics, 477 Madison Avenue, New York, 10022, U.S.
}%
\affiliation{$^{3}$Lund University, Department of Astronomy and Theoretical Physics, Lund, Sweden
}
\affiliation{$^{4}$Technion - Israel Institute of Technology, Physics Department, Haifa, Israel 32000
}

\date{\today}

\begin{abstract}
All scientific claims of gravitational wave discovery to date rely on the offline statistical analysis of candidate observations in order to quantify significance relative to background processes. The current foundation in such offline detection pipelines in experiments at LIGO is the matched-filter algorithm, which produces a signal-to-noise-ratio-based statistic for ranking candidate observations. Existing deep-learning-based attempts to detect gravitational waves, which have shown promise in both signal sensitivity and computational efficiency, output probability scores. However, probability scores are not easily integrated into discovery workflows, limiting the use of deep learning thus far to non-discovery-oriented applications. In this paper, the Deep Learning Signal-to-Noise Ratio (\deepsnr) detection pipeline, which uses a novel method for generating a signal-to-noise ratio ranking statistic from deep learning classifiers, is introduced, providing the first foundation for the use of deep learning algorithms in discovery-oriented pipelines.
The performance of \deepsnr~is demonstrated by identifying binary black hole merger candidates versus noise sources in open LIGO data from the first observation run. High-fidelity simulations of the LIGO detector responses are used to present the first sensitivity estimates of deep learning models in terms of physical observables. The robustness of \deepsnr~under various experimental considerations is also investigated. The results pave the way for \deepsnr~to be used in the scientific discovery of gravitational waves and rare signals in broader contexts, potentially enabling the detection of fainter signals and never-before-observed phenomena.

\end{abstract}

\maketitle

\section{Introduction}
\label{sec:Intro}

In recent years, modern machine learning (ML) techniques, also known as deep learning, have attracted considerable interest in the study of gravitational waves (GWs), particularly for the identification of compact binary mergers~\cite{huertadeepfilter,firstbbhprl,huertabbh-id-param,beijinggroup,harvardbns-id,harvardbns-id-param,ligomlrev,gwskynet,sigmanet,magicbullet}. The more recent of these studies~\cite{sigmanet} have demonstrated the significantly better sensitivity of ML classifiers at mitigating the impact of transient non-Gaussian noise artifacts (glitches) versus matched-filter (MF)-based algorithms.

With GW detectors like LIGO~\cite{ligodet}, glitches are the dominant noise source affecting detection pipelines, all of which are based on the MF algorithm~\cite{pycbc,gstlal}. Because of this, GW-detection pipelines typically require extensive glitch mitigation~\cite{ligoglitches,ligoglitches2}, as current GW discoveries~\cite{gwtc1,gwtc2,gwtc3} would not have been possible without them.

As many of the ML studies above have noted, the MF algorithm is also computationally expensive. In an MF-based workflow or \textit{scan}, each slice of LIGO strain data must be compared against a large multidimensional GW \textit{template bank} representing waveforms described by all of the possible parameters that determine the dynamics of a GW source. In practice, it is infeasible to probe the full parameter space that describes the dynamics of most GW sources, which further reduces the ultimate sensitivity achievable with an MF-based workflow. By contrast, in an ML-based workflow, each slice of the strain data needs to be evaluated only once, making it feasible to train the ML classifier on the full GW parameter space and run the classifier in real time. Real-world applications of ML-based GW classification have thus centered on
real-time or \textit{online} detection to enable, for instance, multimessenger astronomy~\cite{huertadeepfilter,harvardbns-id,mitrealtime,huertahpc}. 

All scientific claims of GW discovery~\cite{gwtc1,gwtc2,gwtc3}, however, are based on \textit{offline} GW-detection pipelines~\cite{pycbc,gstlal}. These pipelines entail more careful statistical analysis than is feasible online, in order to properly quantify the significance of any candidate GW observations. The statistical significance of an observation is expressed in terms of the false alarm probability (FAP or $p$-value) which is the probability of making the observation assuming it was due to some glitch-like background process, or its equivalent rate, the false alarm rate (FAR). A major component of an offline detection pipeline is thus a rigorous estimation of contributing background processes, which typically requires the analysis of a much larger data sample beyond the candidate observation itself. Because of these differences in priorities---rapid response for online detection and rigorous significance quantification for offline detection---the two types of pipelines operate largely independently of one another. Moreover, most of the promising ML results in GW detection are only usable for online detection in their current form because they do not offer a framework for significance quantification with the same level of rigor as MF-based detection pipelines~\cite{ligomlrev,magicbullet}. This is largely because current GW ML classifiers output a probability score, whereas offline detection pipelines require a signal-to-noise ratio (SNR)-type detection statistic that only the MF algorithm provides to date. An SNR statistic tends to separate the distribution of signal-like outliers from those of noise sources, making a distribution in the SNR statistic straightforward to statistically analyze for significance quantification. On the other hand, the probability score of an ML classifier tends to compress the distribution of signal-like outliers, thereby obscuring their separation from a distribution of noise sources. Although this does not make statistical analysis of a probability score distribution impossible, it makes analysis highly cumbersome. Moreover, because of finite float precision, the compression in the distribution of the probability score can cause GW candidates with very high probability scores to be truncated, making their statistical analysis impossible~\cite{mllossbit}.

To address the above issues, we use the raw output of an ML classifier to define a new detection statistic that is normalized to give an SNR interpretation, making integration with existing offline detection pipelines straightforward. We demonstrate how this detection statistic is used to build an ML-based pipeline for identifying GW signal candidates in an offline-oriented workflow, which we call \deepsnr. A broader strategy for developing a fully ML-based offline detection pipeline that uses \deepsnr~as a backbone and does not require the MF algorithm is also described.

In addition to allowing ML classifiers to be used for offline GW detection for the first time, \deepsnr~introduces a number of improvements over previous GW ML classifiers that push the state of the art in detection sensitivity and experimental fidelity and validation. Specifically, \deepsnr~introduces the first use of multidetector waveforms that incorporate multiparameter GWs simulated with high-fidelity detector response. This allows \deepsnr~to learn differences in GW source distribution and thus detector arrival time. Additionally, \deepsnr~makes novel use of a \textsc{ResNet} convolutional neural network (CNN) for GW classification. As an illustrative example, \deepsnr~is trained to distinguish binary black hole (BBH) mergers from background noise sources found in LIGO data.

To validate \deepsnr, we analyze real and simulated BBH mergers in the LIGO detectors. Open data~\cite{gwosc} collected from both the LIGO-Hanford (H1) and LIGO-Livingston (L1) detectors during the O1 data-taking period is used. The physics reach of \deepsnr~is parametrized in terms of key BBH observables to illuminate the pipeline's ability to combine information from LIGO detectors with disparate angular responses. The detection sensitivity of \deepsnr~is compared with the MF algorithm in a simplified benchmark that applies no post-processing to remove glitches. \deepsnr~is able to detect BBH signals with orders-of-magnitude-better sensitivity, demonstrating its ability to perform out-of-the-box glitch mitigation, without which MF-based GW detection would not be possible.

Our ML-versus-MF benchmark complements those found in the literature. The earliest of these benchmarks~\cite{huertadeepfilter,firstbbhprl} were performed using time-domain samples with simulated Gaussian noise and simulated, nonspinning BBH mergers at a fixed sky position. These were followed by studies utilizing noise from real data taken a few seconds within known GW discoveries~\cite{huertabbh-id-param} or from glitches identified by the \textsc{GravitySpy} project~\cite{sigmanet}. The latter of these studies simulated nonspinning BBH sources from various sky positions, applied glitch mitigation, and used multidetector inputs from the time-versus-frequency domain. These studies, however, did not account for differences in detector arrival time. Our benchmark utilizes noise taken from several hours' worth of real data, including any and all glitches therein,
and multidetector inputs from the time domain without any glitch mitigation. We find that ML classifiers trained on time-domain instead of time-versus-frequency-domain inputs yield significantly more sensitive detection. In addition, we use BBH signals simulated over a much broader range of parameters and accurately model detector response to capture differences in detector arrival times.

The robustness of \deepsnr~is investigated in a number of first studies that quantify the dependence of an ML-based GW classifier on various experimental considerations, including the number of LIGO detectors, detector evolution, and simulation modeling. It is our hope that these studies pave the way for the first integration of ML-based models into offline GW-detection pipelines, potentially allowing challenging and never-before-seen GW signals such as binary neutron star (BNS) mergers to be detected.

This paper is organized as follows. Section~\ref{sec:Data} outlines the LIGO data sets we used, and Section~\ref{sec:Process} describes how they are processed for analysis. In Section~\ref{sec:Method}, we detail the \deepsnr~methodology for performing ML-based signal identification with significance quantification. The results of demonstrating and validating \deepsnr~in the context of BBH identification are presented in Section~\ref{sec:BBH}. A discussion of how our results fit into the broader goal of developing a real-world, fully ML-based discovery pipeline is given in Section~\ref{sec:Discussion}. A summary of our results is provided in Section~\ref{sec:Summary}.

\section{Data samples}
\label{sec:Data}

Data samples from the Gravitational Wave Open Science Center (GWOSC) open LIGO data set, collected during the first observation period (O1), are used~\cite{gwosc}. We use the strain data from both available GW detectors during O1 data-taking, H1 and L1, sampled at 4 kHz. For all time segments used for analysis, we require that both detectors were online and collecting data. Additionally, the data quality flags in both detectors must pass the highest-quality flag, \texttt{CAT3}, for both GW coalescence and burst categories~\cite{gwosc}. We emphasize that these requirements do not guarantee that the selected data will be free of glitches~\cite{ligoglitches,ligoglitches2}.

For the purposes of this paper, we require only a subset of the O1 data. Separate time segments passing the above requirements are selected for the training, validation, and testing of our ML pipeline, as listed in Table~\ref{tab:datasets}. The training and validation sets roughly correspond to earlier parts of O1 data-taking and are used to train and optimize the ML model, respectively. The test sets consist of samples roughly collected from the start, middle, and end of O1 data-taking. All time segments are chosen to have the same duration and do not contain any of the discovered O1 GW events~\cite{gwtc1}.

\begin{table}[!htbp]
\caption{List of GWOSC data sets used for study, corresponding to data collected by the H1 and L1 detectors during the O1 data-taking period.}
\label{tab:datasets}
\begin{ruledtabular}
\begin{tabular}{lccc}
 Data set    & Start time  & Duration & Entries \\
             & (GPS time)  & (hours)  & ($10^3$)\\
 \hline
 Training    &  1126297600 & 6.6      & 48\\
 Validation  &  1126400000 & 4.4      & 32\\
 Test, start &  1126785024 & 12.1     & 87\\
 Test, mid   &  1131302912 & 12.1     & 87\\
 Test, end   &  1136578560 & 12.1     & 87\\
 \hline
\end{tabular}
\end{ruledtabular}
\end{table}

We assume that the GWOSC data sets do not contain any GW signals; waveforms that resemble GWs are strictly interpreted as the results of glitches. The above samples thus define our \textit{background} sample, which is derived entirely from data. Compared with previous studies with a data-driven background sample~\cite{huertabbh-id-param,harvardbns-id-param}, our sample is larger by an order of magnitude and thus contains a correspondingly larger population of glitches. This is particularly relevant because glitches are the dominant background contaminant in the regimes where GW signals can be detected. To prepare a data sample containing BBH signals, simulated pure GW signal waveforms are generated. The BBH signal waveforms are generated by \textsc{PyCBC}~\cite{pycbcsw} using the \textsc{SEOBNRv4} approximation~\cite{SEOBNRv4}. The black hole pair interacting in the BBH merger has constituent masses uniformly distributed in $m_1,m_2 = 10$--50 $\msun$ and uniformly distributed spin perpendicular to the merger plane. The coalescence, inclination, and polarization of the merger are also sampled from uniform distributions of the respective parameter domains to cover the parameter space of possible mergers. Relative to the LIGO detectors, the distance to the BBH merger (luminosity distance $\dL$) is uniformly sampled from a volume with radius $\dL =$ 100--500 Mpc, and the sky position is uniformly sampled in right ascension and declination~\footnote{Note that this choice of sky position sampling will tend to oversample the polar regions relative to a sampling that is uniform in solid angle.}. The generated BBH signal is then simulated through the respective H1 or L1 detector response. The detector response is modeled in \textsc{PyCBC} by projecting the GW signal onto the LIGO detector antennas, taking into account the relative sky position of the GW source, the GW times of flight, and the rotation of the Earth at the time of simulated detection~\cite{pycbcsw}. Importantly, this means the relative phase and time delay between the simulated BBH signals detected in the H1 and L1 detectors are being accounted for. The resulting BBH waveform is overlaid onto the background sample waveforms at random positions. The combined data define our BBH \textit{signal} sample. We emphasize that only the pure GW signal is derived from simulation---the noise component of the waveforms is taken directly from data. Notably, the fully simulated sky position and detector response in our signal sample enables us to study these effects for the first time in a ML study.

\section{Data processing}
\label{sec:Process}

The waveforms in both signal and background samples are first whitened by normalizing the power spectral density over the LIGO frequency range~\cite{ligosgextract}. To remove potentially biased uncalibrated frequencies, the whitened waveforms are then processed through a band-pass filter between 16 Hz and 2 kHz ~\cite{ligosgextract,gwosc}. The resulting waveforms are then processed through a notch filter at 60, 120, and 180 Hz to remove resonances associated with the detector power supplies~\cite{ligosgextract,gwpysw}. We apply a second band-pass filter between 36 and 320 Hz to select only frequencies where the power spectral density in the BBH signal is expected to be above that of detector noise~\cite{gwoscws}. As discussed in Appendix~\ref{app:trobust}, we find this choice of band-pass filter range to be close to optimal. Finally, the resulting strain values are scaled by a sample-wide constant so that the sample-wide root-mean-square (RMS) is approximately unity.

We use only the native time-domain waveform strains in our analysis. As investigated in Appendix~\ref{app:tvf},
the use of time-domain inputs results in a modest sensitivity advantage over frequency-domain inputs. To convert the continuous waveform strains to fixed-length arrays suitable for ML training, we apply a sliding window over the selected and processed time segments described earlier in Section~\ref{sec:Data}. We use a window of duration 1~s with a stride of 0.5~s. Since the strain data we use is sampled at $2^{12} \approx 4$ kHz, each entry in the output training sample will be an array of length 4096. The choice of stride length ensures that every point in the source waveform appears twice in the training sample (or once if near a segment boundary). This guarantees that at least one of the two extracted waveforms will not be near the edge of the array. For the simulated GW signals that are overlaid on data, we ensure that the signal is positioned no closer than 0.2~s (0.05~s) to the left (right) edge of the 1~s time window.

The total number of samples selected in each data set is given in Table~\ref{tab:datasets} (rightmost column). Our results are not overly sensitive to the time duration of the sliding window, as shown in Appendix~\ref{app:trobust}. We expect this to be the case for as long as the duration over which the GW signal is above the detector noise threshold is much less than the window duration.

\section{Signal identification methodology}
\label{sec:Method}

In this section, we describe the methodology for building the \deepsnr~pipeline. First, we describe the construction and training of the ML classifier that serves as its backbone. Using the output of the ML classifier, we introduce the detection statistic that will be used for ranking GW signal candidates. We then describe the procedure for identifying signal candidates based on detection significance. Finally, we outline our strategy for validating the above pipeline.

\subsection{ML classifier}
A \textsc{ResNet} model, a type of CNN, forms the basis of our ML-based classifier. As many have already observed~\cite{huertadeepfilter,firstbbhprl,magicbullet}, CNNs are well suited to extracting signals from waveforms because of the translational symmetry of GW signals in waveform data. The addition of the residual blocks found in \textsc{ResNet} models helps overcome the performance degradation associated with increased ML model depth, as demonstrated in image classification tasks~\cite{resnet}. The details of the ML classifier architecture are deferred to Appendix~\ref{app:arch}. The inputs to the classifier are the time-domain strain data converted to fixed-length arrays (Section~\ref{sec:Process}). The strain data from both H1 and L1 are combined to form a single, multichannel array of shape $(2, 4096)$ corresponding to a single input to the classifier. In contrast to single-detector inputs, multidetector inputs allow the classifier to better exploit independent information between different detectors, as noted in other multidetector experiments~\cite{qvg_e2e}. The ML classifier is then trained to distinguish the BBH signal from the background sample (described in Sec~\ref{sec:Data}) by minimizing the binary cross-entropy loss between the two sample types using the \textsc{ADAM} optimizer. We train for 50 epochs but select the model from the epoch with the best performance, as defined in Section~\ref{sec:Method:valid}. We make no attempt to stagger the supply of training samples over time, unlike the curriculum learning used in Refs.~\cite{huertabbh-id-param,mllossbit}. The raw output of the classifier, known as the logit $l$, represents the classification score. As a matter of convention, the classifier is trained to output larger logit values for samples that are more signal-like.

\subsection{Detection statistic}
\label{sec:Method:qstat}
For many ML classification applications, including those targeting online GW detection, one typically converts the raw logit value to a normalized probability score by applying the sigmoid function, $\mathrm{sigmoid}(l) = [1+e^{-l}]^{-1}$~\cite{mlref}. A probability threshold is then defined, above which the sample is categorized as a GW signal candidate, otherwise as a background candidate.

LIGO searches for GWs require some form of significance quantification, typically obtained using a statistical hypothesis test. In such tests, a potential signal candidate is ranked against background samples based on a detection statistic (also known as the test statistic). The significance of the signal candidate is quantified by the FAP, which is the probability for a background sample to rank with a detection statistic higher than that observed for the signal candidate. Claims of scientific evidence conventionally require a measured FAP $\lesssim 1.3\times10^{-3}$, corresponding to a significance of at least $3\sigma$.

\begin{figure}
\centering
\includegraphics[width=\linewidth]{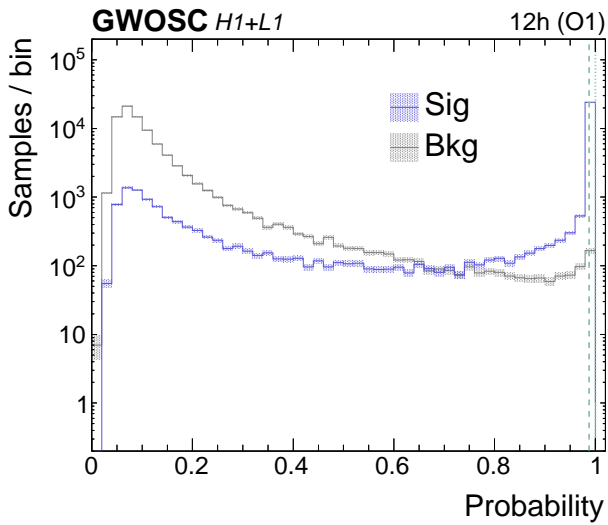}\\
\caption{Representative probability score distributions for signal (Sig, blue histogram) and background (Bkg, gray histogram) samples evaluated using the ML classifier trained on H1+L1 inputs. The statistical uncertainties are shown as the filled bands around the histograms. The green vertical lines indicate the working points of 3 (dashed) and $4\sigma$ (dotted) significance, as defined in Section~\ref{sec:Method:cands}. The separation between signal and background samples at these high-significance levels is not apparent, making statistical analysis difficult.}
\label{fig:prob}
\end{figure}

Although the probability score described above can function as a detection statistic, it is cumbersome for GW detection where one is primarily interested in signal-like distribution outliers. This is because the sigmoid function tends to compress the probability scores of signal-like outliers near its maximum range of 1. As illustrated in Fig.~\ref{fig:prob}, this tends to obscure the separation between signal and background samples. Moreover, finite float precision can potentially cause large probability scores to be truncated and rounded off to 1, making it impossible to rank samples beyond a certain significance level. Increasing the float precision, as proposed in Ref.~\cite{mllossbit}, mitigates the last of the concerns but does not address the fundamental limitations of the probability score-based formulation. Indeed, this behavior is responsible for a similar problem in ML known as the vanishing gradient problem. To better analyze the distribution of signal-like outliers for the purpose of significance quantification, we propose an alternate detection statistic that does not saturate at large $l$. To preserve the ranking of samples, we require that this detection statistic be a monotonically increasing function of $l$. Although the raw logits themselves satisfy the above requirements, they are not positive-definite, complicating their interpretation. We therefore propose the following function of the logits as a detection statistic,
\begin{equation}
    \qstat(l) = \frac{1}{N}\mathrm{ln}(1+e^{l}),\quad -\infty < l < \infty,
\end{equation}
where $N$ is some normalization constant. As $l \rightarrow -\infty$, $\qstat(l) \rightarrow e^{l} > 0$ and as $l \rightarrow \infty$, $\qstat(l) \rightarrow l$, satisfying the desired criteria. The choice of functional form fulfilling the same requirements is not unique. Indeed, some of the functions that address the vanishing gradient problem can also be used. We fix $N$ such that the RMS of the detection statistic over some reference background sample is unity, $\langle \qstat \rangle_{\mathrm{noise}} = 1$. With this choice of normalization, $\qstat$ carries an SNR interpretation, similar to existing MF-based LIGO detection statistics. Note, however, that there is no reason, \textit{a priori}, that $\qstat$ should scale with the FAP at the same rate that other LIGO SNR-based detection statistics do. A distribution of $\qstat$ that highlights the separation between signal and background samples ranked with high significance is shown in Fig.~\ref{fig:nnout} (top).

\subsection{Signal candidates}
\label{sec:Method:cands}
A set of GW signal candidates is identified by requiring a minimum value of $\qstat$.
The corresponding detection significance of a signal candidate with value $\qstat$ is then measured by the FAP with respect to some background sample. In our analysis, we define the FAP as the fraction of background samples ranked above a given $\qstat$ threshold. In practice, a FAP working point is first chosen, then the associated $\qstat$ threshold is calculated. We choose a FAP working point corresponding to a significance of 3$\sigma$, or about FAP $\lesssim 1.3\times10^{-3}$. This defines our pipeline for identifying GW signal candidates with significance quantification.

The above definition for the FAP is standard for binary classification tasks~\cite{mlref} but differs from the FAR conventions used in LIGO analyses which are based on observation time~\cite{ligosgextract}. Moreover, the background sample we use to measure the FAP also differs from those used in LIGO analyses to measure the FAR. In real-world LIGO searches, the background sample used is obtained either from time-shifted data samples subject to waveform compatibility requirements~\cite{pycbc,findchirp} or modeled from the noise evolution~\cite{gstlal,gstlalbkg}. In our simplified analysis, all data-only samples are defined to be background samples. Thus, the FAP values we present are not directly comparable to those presented in LIGO discoveries. In order to perform real-world background estimation with \deepsnr, techniques beyond those developed in this paper are required. These are discussed in Section~\ref{sec:Discussion}.

\subsection{Validation strategy}
\label{sec:Method:valid}
To validate \deepsnr, we demonstrate its use in open LIGO data (Section~\ref{sec:Data}). We quantify the expected BBH detection sensitivity using the true alarm probability (TAP) at various FAP working points. The TAP is defined as the fraction of signal samples whose $\qstat$ values are ranked above some threshold specified by a FAP working point. For the chosen FAP working point with $3\sigma$ significance, we provide estimates of the expected physics reach with respect to various BBH observables of interest.

A number of studies are presented to characterize the dependence of \deepsnr~on various experimental quantities of interest. First, we study the impact of mutual detector information. We compare the signal sensitivity using the nominal H1+L1-trained ML classifier with those using single-detector-trained ML classifiers. To benchmark the ML results, we compare them with their MF-based counterparts. A simplified MF algorithm is run~\cite{gwosc,gwoscws}. For each signal sample, only the overlap with its own GW signal template is calculated. After normalizing to the same time window in the background sample (i.e., without the GW signal overlaid), the value at the point of peak overlap is taken. This defines the MF-based SNR detection statistic for signal samples. For background samples, the SNR with respect to a randomly generated GW signal template is used. The signal template is generated from the same parameter space as the one used for producing the signal samples (described in Section~\ref{sec:Data}). This will tend to underestimate the peak SNR of the background sample and thus improve the separation between the signal and background samples in the MF benchmark. In real-world use, when the MF scan is performed over the full signal template bank, templates yielding higher SNRs are more likely to be selected. In addition, for computational reasons, real-world MF scans typically use templates parametrized by only the BBH masses and spins~\cite{gwtc1}, not the broader set that we use. As others have noted~\cite{huertabbh-id-param,harvardbns-id-param}, this is an important computational advantage that the ML-based approach holds over the MF one, and it has potential implications for detection sensitivity over the full BBH parameter space. The post-processing steps used by LIGO analyses to mitigate the effect of glitches, such as the $\chi^2$ re-weighting~\cite{pycbc,chi2}, are not performed in this basic benchmark.

Second, we study the impact of the time evolution of the LIGO detectors. We compare the $\qstat$ distribution in different O1 time segments, for both signal and background samples. This additionally serves as a check of (a) the ML classifier's behavior under noise conditions different from those used to train it and (b) its ability to exploit improvements in noise mitigation.

Lastly, we study the impact of simulation modeling. We compare the compatibility of the $\qstat$ distributions between actual BBH signals in data versus those simulated with matching BBH parameters at the 90\% confidence interval (CI).

\section{Results}
\label{sec:BBH}

\begin{figure}[!htbp]
\centering
\includegraphics[width=\linewidth]{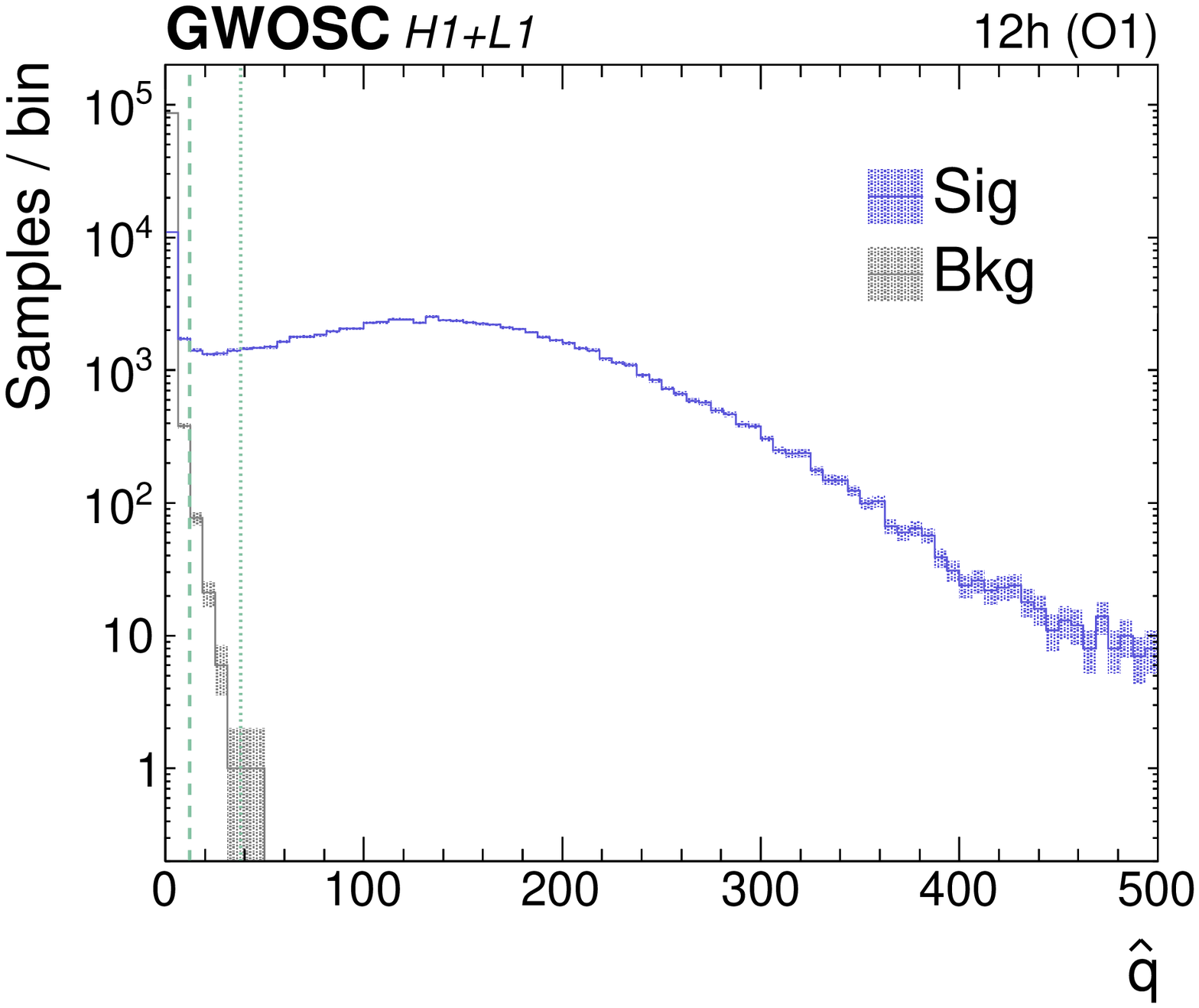}\\
\hspace{10pt}\includegraphics[width=.91\linewidth]{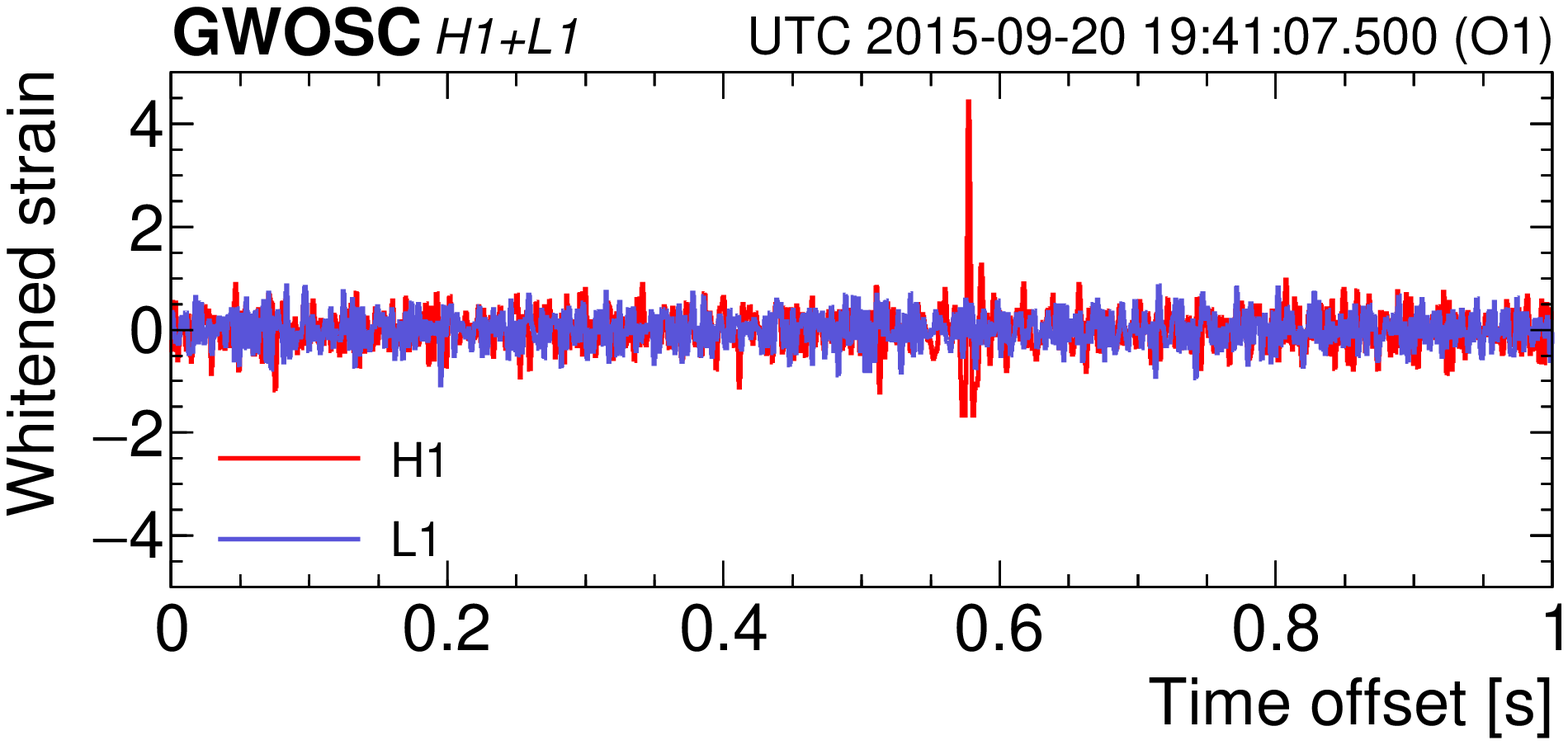}\\
\hspace{10pt}\includegraphics[width=.91\linewidth]{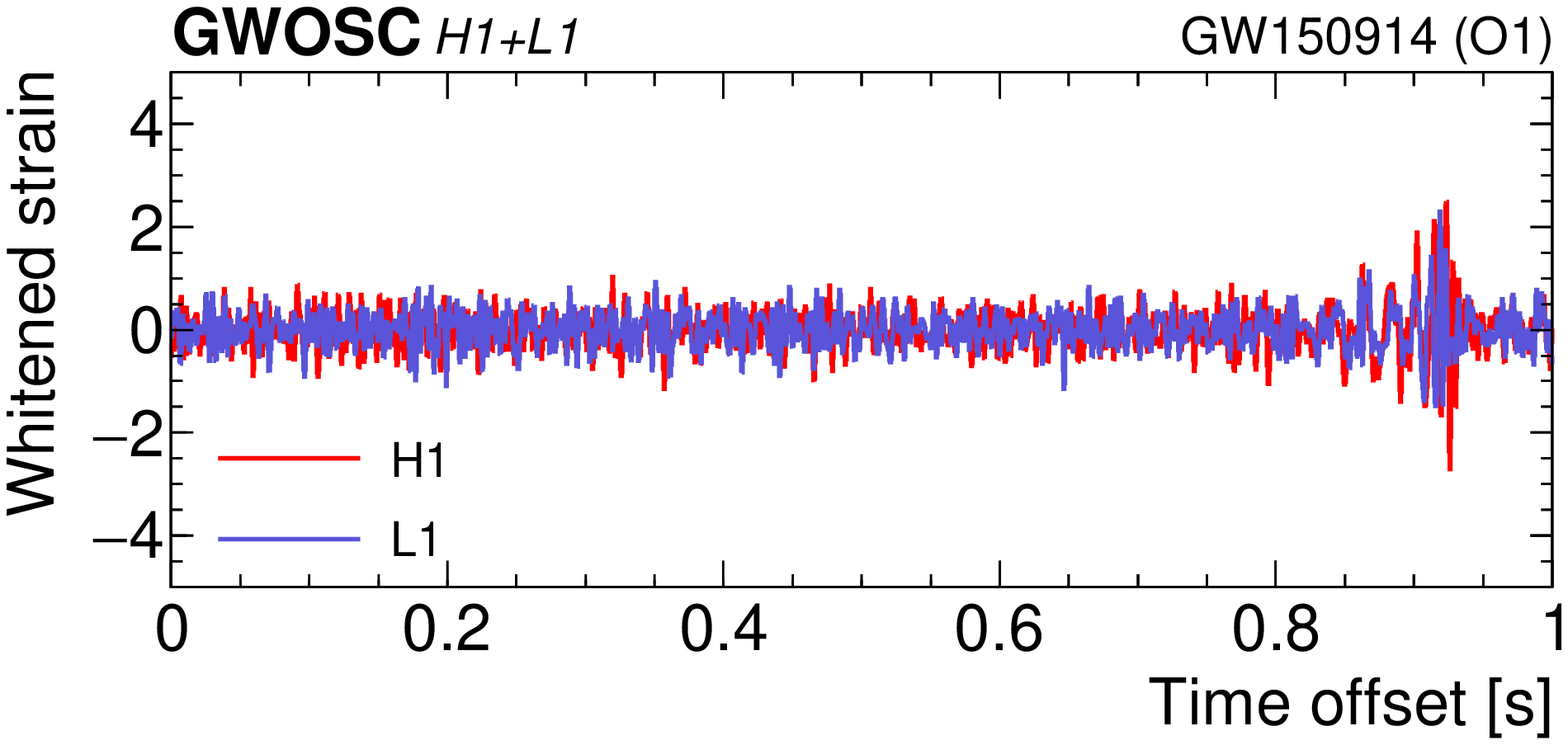}
\caption{Top: Distributions of the detection statistic $\qstat$ for signal (Sig, blue histogram) and background (Bkg, gray histogram) samples evaluated using the ML classifier trained on H1+L1 inputs. Both distributions are normalized to the first 1000~s of the background sample to give an SNR interpretation. Note that the scale and range of the $\qstat$ values are not directly comparable to those of LIGO MF-based detection statistics, as discussed in Section~\ref{sec:Method:qstat}. The statistical uncertainties are given by the filled bands around the histograms. The green vertical lines indicate the working points of 3 (dashed) and $4\sigma$ (dotted) significance. Clear separation between the signal and background samples is visible above the $4\sigma$ significance level, beyond which the signal distribution starts to peak. Middle and bottom: Representative whitened waveforms measured at H1 (red) and L1 (blue) for a background sample detected with a significance above $4\sigma$ (middle) and for \textsc{GW150914} (bottom) for scale.}
\label{fig:nnout}
\end{figure}

The distribution of the $\qstat$ detection statistic for samples obtained from the early part of O1 data-taking is shown in Fig.~\ref{fig:nnout} (top). The samples are evaluated using our ML classifier trained on inputs corresponding to combined H1+L1 strain data (described in Section~\ref{sec:Process}. The distributions for both the signal and background samples are normalized to the first 1000~s of the background sample to give an SNR interpretation. Compared to the probability score-based distribution in Fig.~\ref{fig:prob}, the SNR-based distribution in $\qstat$ clearly highlights the separation between signal and background samples. Signal samples populating the $\qstat<1$ region are primarily more distant, lower-mass BBH mergers with measured strains below that of detector noise. Signal samples in the far right-side tail of the distribution are dominated by closer-by, higher-mass mergers. Even though the duration over which larger-mass merger signals are above the detector noise is shorter, the greater energy they release produces correspondingly louder GW signals that are easier to detect. Signal samples in the intermediate bulk of the distribution represent a smooth continuum of BBH masses and luminosity distances. The signal distribution peaks just after the background distribution tapers off. Background samples above the $4\sigma$ significance level (dotted line) consist mainly of waveforms where a burst-like glitch occurred in one detector but not the other, as illustrated in Fig.~\ref{fig:nnout} (middle). These can resemble a GW signal that was well detected in one detector but not the other---for instance, if the GW came from a sky position where one of the detectors has weaker sensitivity.

To quantify the expected sensitivity of \deepsnr, we plot in Fig.~\ref{fig:TARvsfprinv} (left) the estimated TAP as a function of various FAP working points. The TAP of the H1+L1-trained ML classifier (black points) ranges between $90\%$ to 70\% for significance levels between 3 to $4\sigma$, respectively. However, note that the exact TAP values will differ depending on the composition of the underlying BBH sample parameter space (outlined in Section~\ref{sec:Data}). The identification of GW signal candidates requires a FAP working point to be chosen. As discussed in Section~\ref{sec:Method:cands}, we choose one corresponding to a detection significance of $3\sigma$. Based on this working point, we can then characterize the expected physics reach or discovery potential of \deepsnr. The estimated TAP as a function of total BBH merger mass $\mtot$ and $\dL$ for this working point is given in Fig.~\ref{fig:phasespace} (top left). BBH signal detection is fully efficient for nearer-by, large-mass mergers and about 40\% efficient for more distant, low-mass mergers, consistent with earlier observations. The expected TAP as a function of sky position for the same working point is plotted in Fig.~\ref{fig:phasespace} (top right). We observe a distinct angular dependence in the signal sensitivity when using the combined H1+L1 ML classifier, which varies in TAP from about 70\% to more than 90\%. The complementarity in angular sensitivity between the two detectors plays an important role in the performance of the combined H1+L1-trained ML classifier, which we discuss in the following section.

\subsection{Mutual detector information}
\label{sec:BBH:multidet}

\begin{figure*}
\includegraphics[width=.45\linewidth]{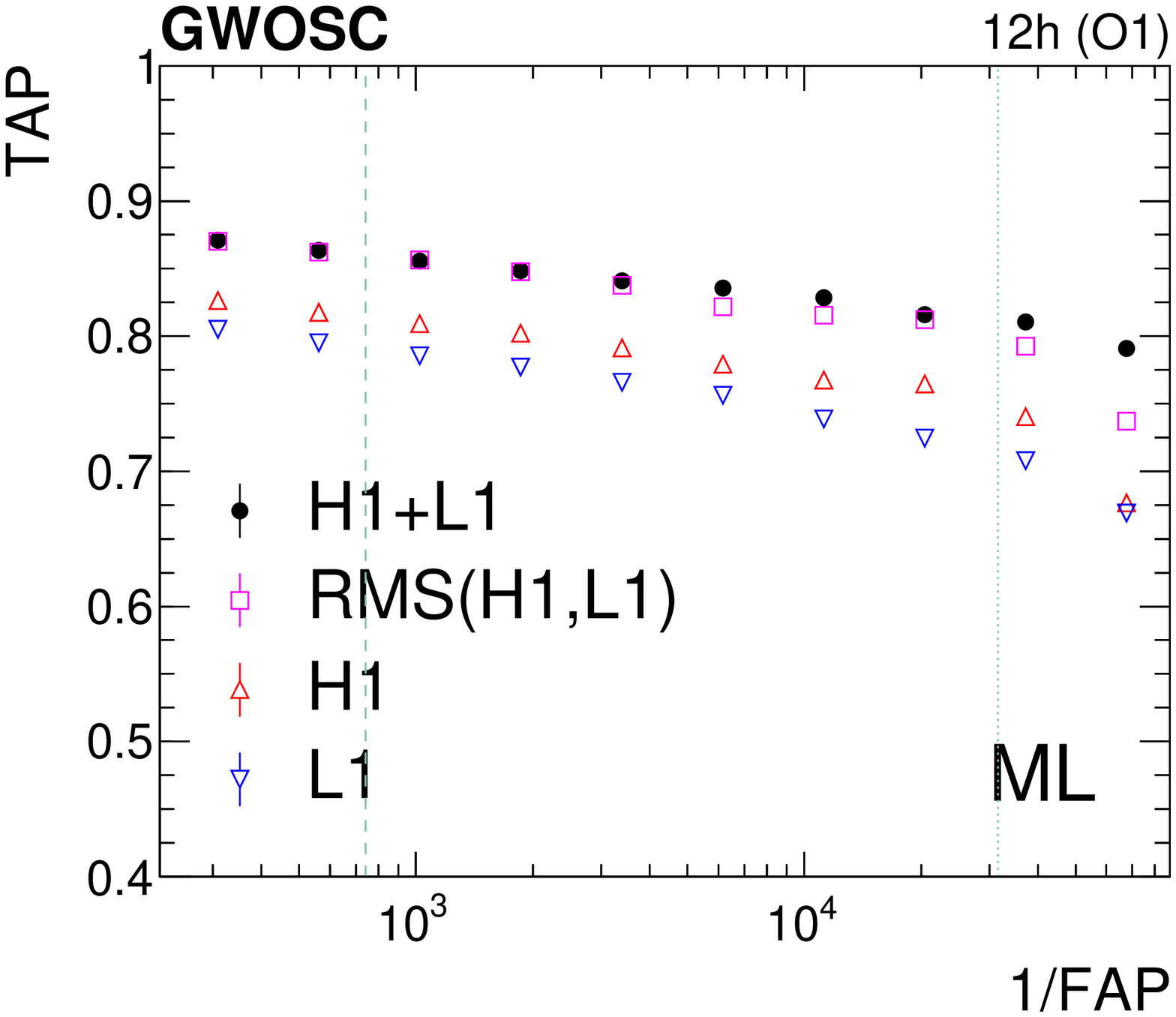}
\hspace{10pt}
\includegraphics[width=.45\linewidth]{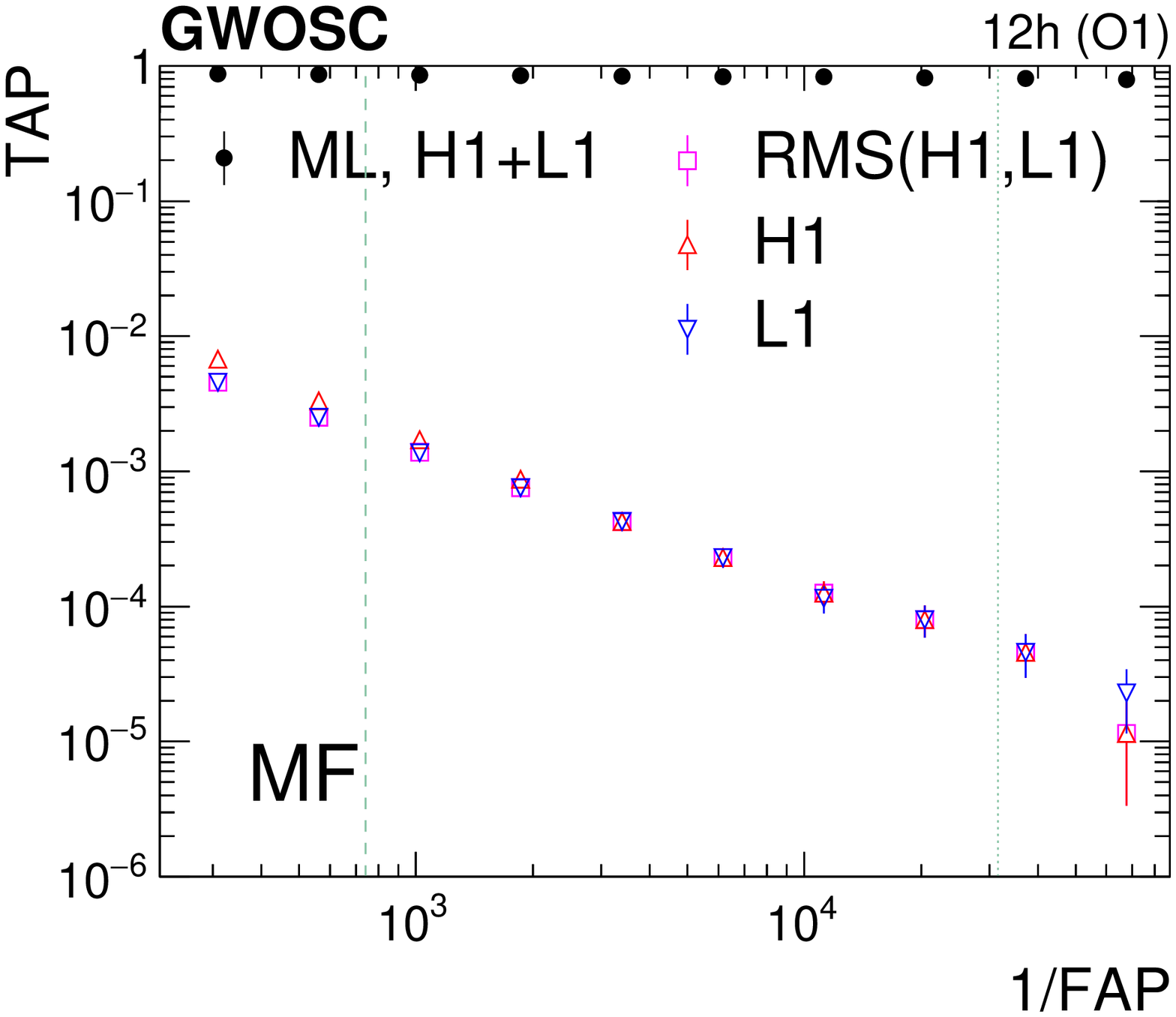}
\caption{Expected TAP as a function of the inverse FAP (1/FAP)\footnote{The choice of inverse FAP on the $x$-axis ensures that, similar to Fig.~\ref{fig:nnout} and the LIGO results in Ref.~\cite{gwtc1}, the significance level increases to the right}. Left: ML-based classifiers trained on inputs from H1+L1 (black points), H1 (right triangles), L1 (blue inverted triangles), and the network combination RMS(H1,L1) (purple squares). Right: MF-based algorithms using inputs from H1 (right triangles), L1 (blue inverted triangles), and their network combination RMS(H1,L1) (purple squares). No post-processing to mitigate the impact of glitches is performed, without which the MF algorithm is known to perform poorly (see text). Glitch removal is expected to improve both the MF and ML results. The results of the H1+L1-trained ML classifier (black points) is shown for comparison. The statistical uncertainties are given by the vertical bars on the markers. Note the difference in the y-axis scales. The green vertical lines indicate the working points of 3 (dashed) and $4\sigma$ significance.}
\label{fig:TARvsfprinv}
\end{figure*}

To assess the impact of mutual detector information on GW signal identification, we compare the sensitivity of \deepsnr~when using the ML classifier trained on H1+L1 against ML classifiers trained on single detectors (1-det). Comparisons of the estimated TAP as a function of different FAP working points are given in Fig.~\ref{fig:TARvsfprinv} (left). We also include the combined network output of the two 1-det classifiers, defined as RMS(H1,L1) = $\sqrt{\qstath^2 + \qstatl^2}$, where $\qstath$ ($\qstatl$) is the $\qstat$ corresponding to the H1- (L1)-trained ML classifier, normalized to the first 1000~s of their respective background samples. We recall from Section~\ref{sec:Method:valid} that the FAP working point is defined by the $\qstat$ threshold where the fraction of background samples ranking higher is equal to the desired FAP value. The corresponding TAP at this FAP working point is thus the fraction of signal samples ranking above the same $\qstat$ threshold. The RMS(H1,L1) classifier has comparable sensitivity to the H1+L1 classifier. Both multidetector classifiers generally outperform either of the 1-det classifiers. The H1+L1 classifier maintains a slight advantage over the RMS(H1,L1) one for significance levels around $4\sigma$ (dotted line). The H1 classifier performs slightly better than the L1 one because of the better noise profile of H1 during O1~\cite{ligoglitches2}. The L1 classifier starts to degrade faster beyond the $4\sigma$ significance level. One computational benefit of the RMS(H1,L1) classifier over the H1+L1 one is that its calculation is factorizable. That is, in analysis applications where one shifts the detector waveforms with respect to one another to derive a background estimate~\cite{pycbc,findchirp}, the classification needs to be run only once for each detector. Afterward, only the pairing of outputs to calculate the RMS needs to be repeated. In the case of the H1+L1 classifier, the same background estimate would require rerunning the classifier for every possible pairing of detector waveforms.

To put the above ML results into context, we perform a benchmark against the MF algorithm. Similar comparisons of 1-det and the networked RMS(H1,L1) performance using the MF algorithm are given in Fig.~\ref{fig:TARvsfprinv} (right). The MF algorithms generally rank with lower TAPs than the ML classifiers by several orders of magnitude. It bears repeating, however, that the MF algorithm \textit{on its own} is known to perform quite poorly in the presence of glitches~\cite{ligoglitches,pycbc}. Indeed, it is only because of painstaking post-processing to remove glitches that any of the BBH discoveries have been possible using an MF-based pipeline. These post-processing steps include down-weighting the SNR statistic based on BBH signal compatibility ($\chi^2$ re-weighting)~\cite{chi2,ligoglitches} and vetoing candidates with incompatible matched templates between detectors, among others~\cite{ligosgextract,pycbc}. Thus, a major strength of the ML-based pipeline is that it achieves high detection sensitivity even without post-processing to remove glitches. Moreover, we expect an ML-based pipeline to \textit{also benefit} from these same glitch-mitigation steps because they are applied independently of the derivation of the SNR detection statistic. However, it is likely that the ML-based pipeline is already exploiting some of the discriminating power derived from these post-processing steps. Real-world gains from \deepsnr~are thus likely to be lower than what we present here. As elaborated later in Section~\ref{sec:Discussion}, additional machinery beyond what is developed in this paper is required to perform a rigorous real-world comparison of the ML- and MF-based pipelines. Our ML results are comparable to those of Ref.~\cite{sigmanet}, although our MF-based results suggest lower TAPs because we do not apply glitch mitigation.

In the current benchmark with no glitch mitigation applied to either the ML- or MF-based pipeline, the addition of a second detector in the MF approach (Fig.~\ref{fig:TARvsfprinv}, right) does not give better detection sensitivity, which differs from the pattern in ML-based detection (Fig.~\ref{fig:TARvsfprinv}, left). The MF-based RMS(H1,L1) performance is ultimately constrained by the weaker of the two detectors. Thus, another advantage of the ML approach is its ability to better exploit correlations across multiple detectors. ML-based GW detection may therefore have an outsized benefit over MF-based detection for large arrays of GW detectors.

\begin{figure*}
\centering
\includegraphics[width=.45\linewidth]{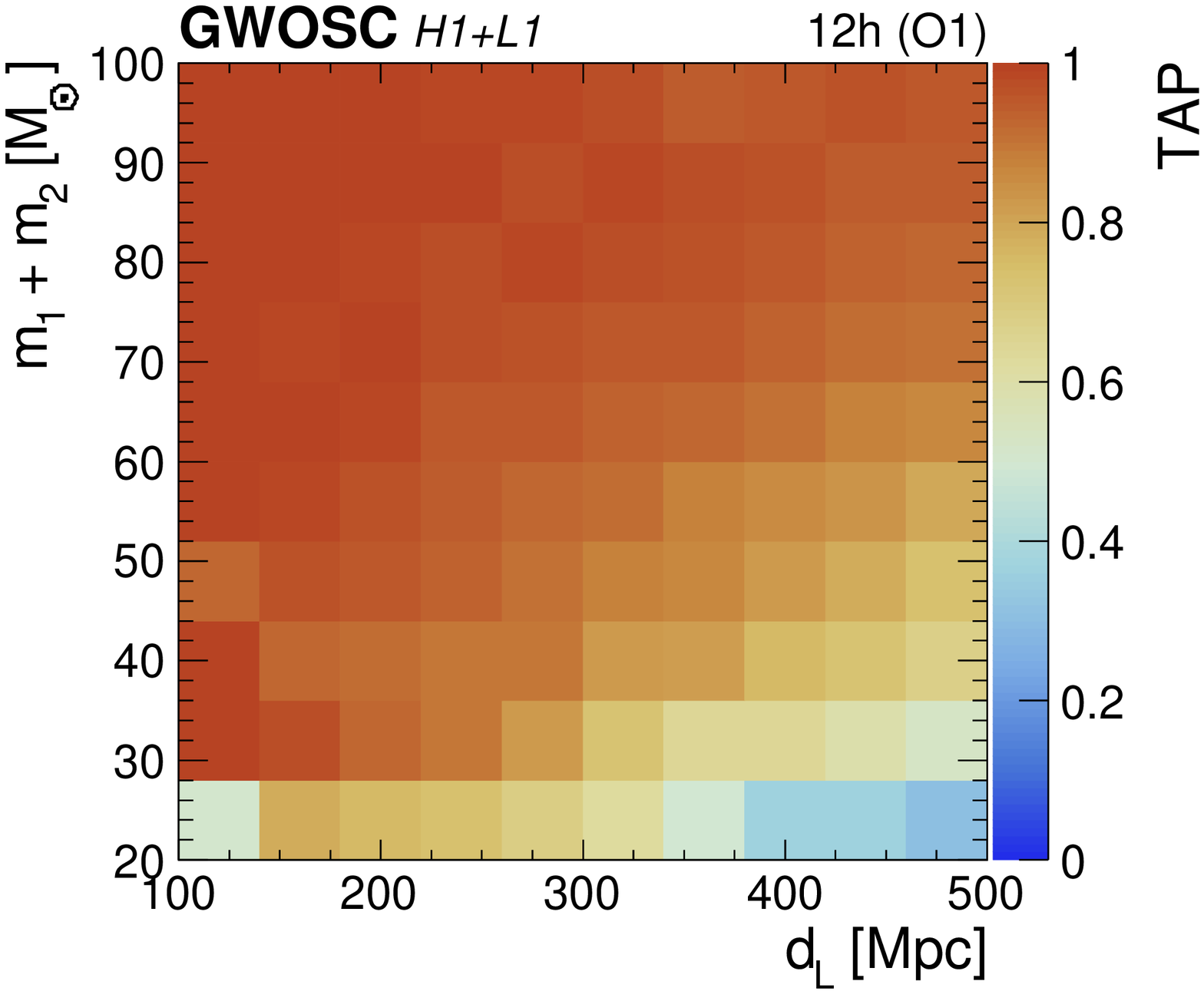}
\hspace{10pt}
\includegraphics[width=.45\linewidth]{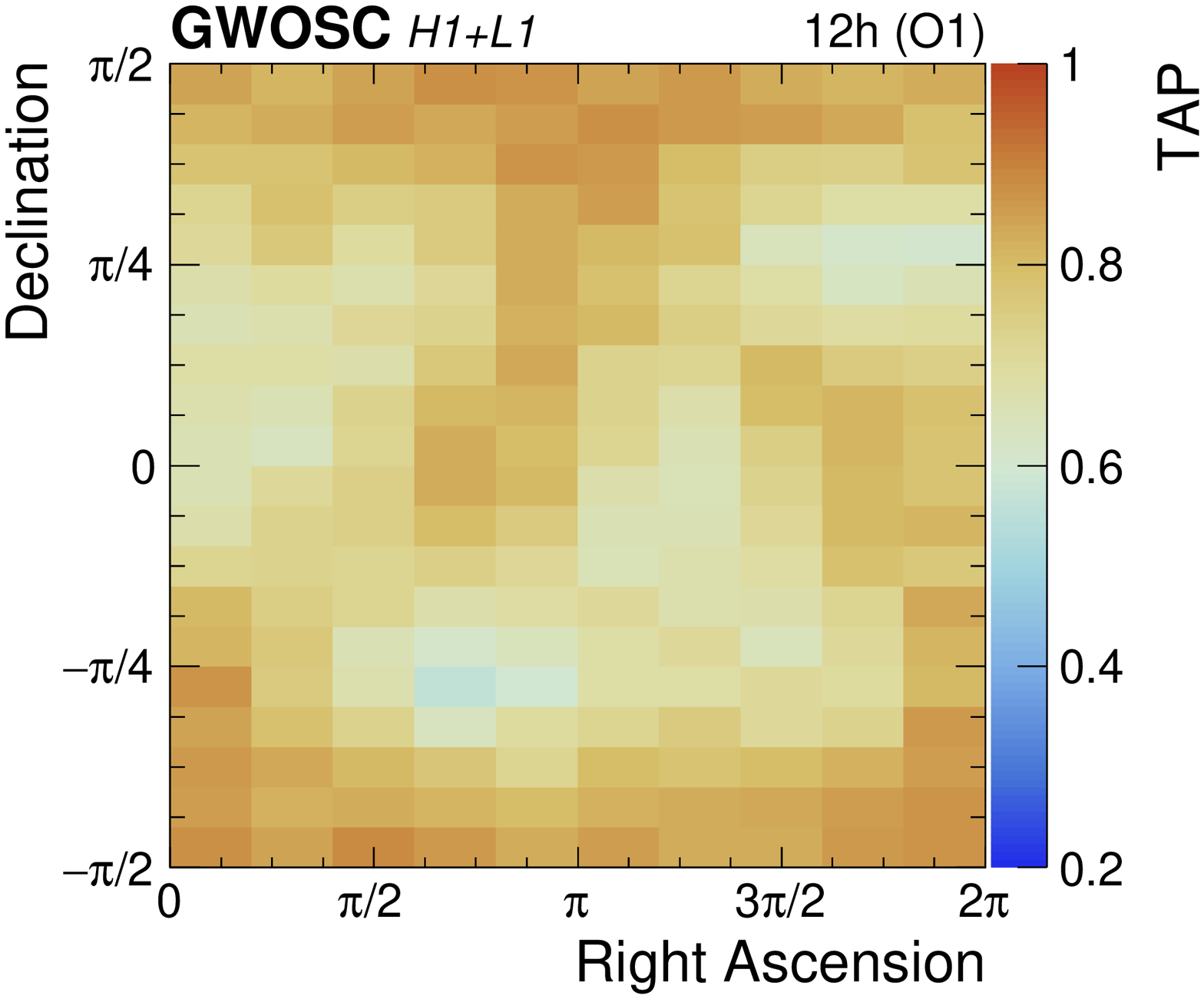}
\\
\includegraphics[width=.45\linewidth]{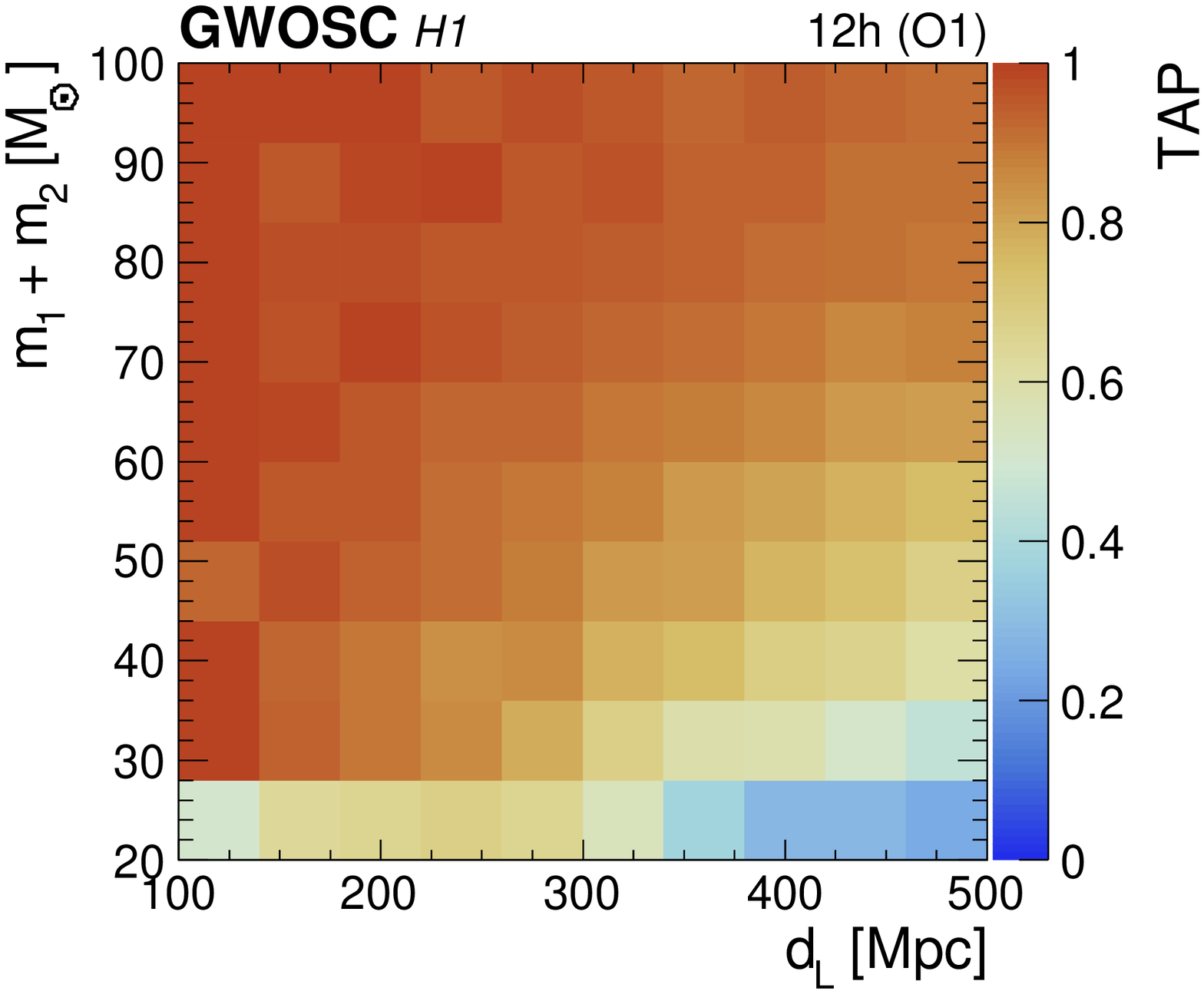}
\hspace{10pt}
\includegraphics[width=.45\linewidth]{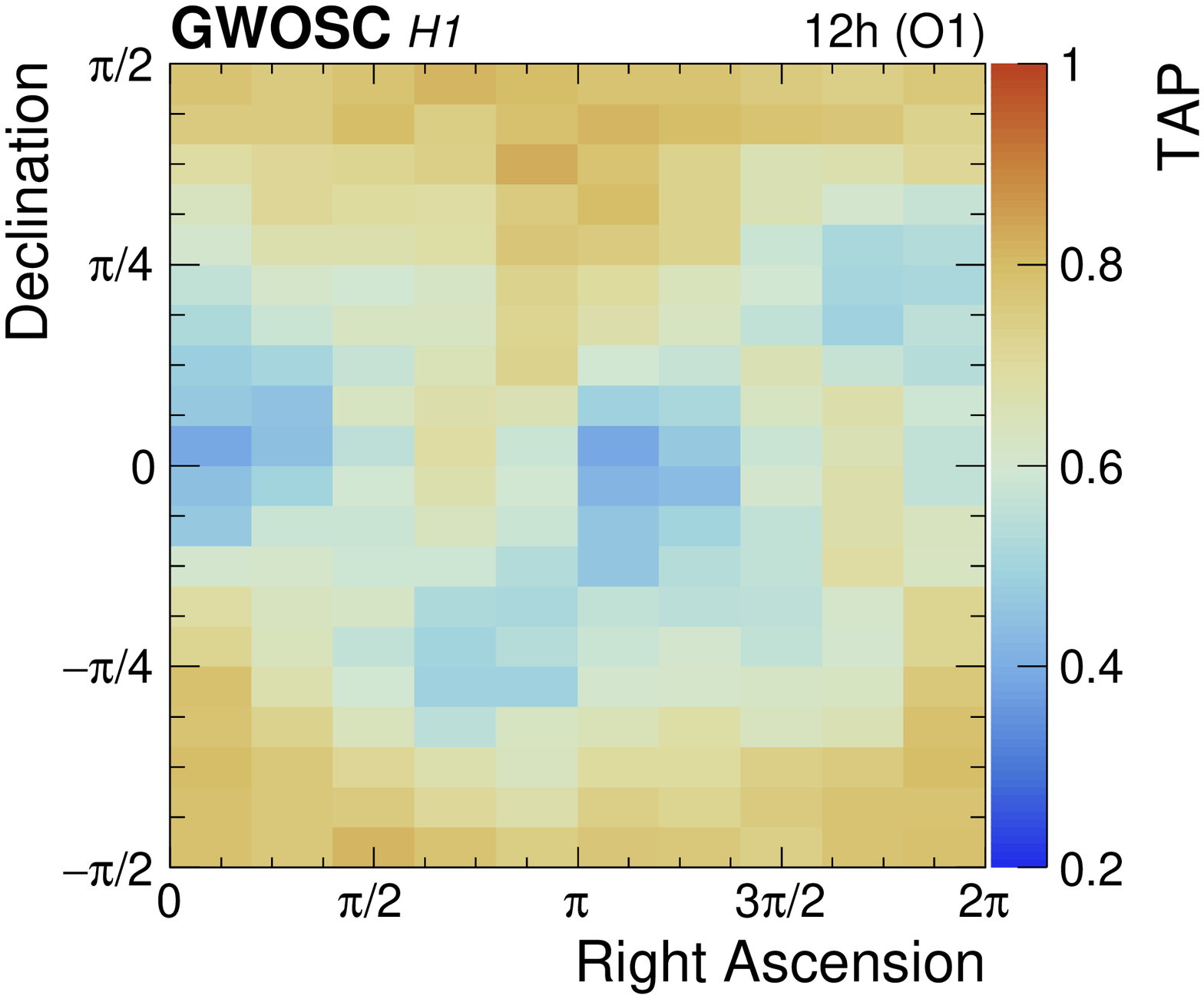}
\\
\includegraphics[width=.45\linewidth]{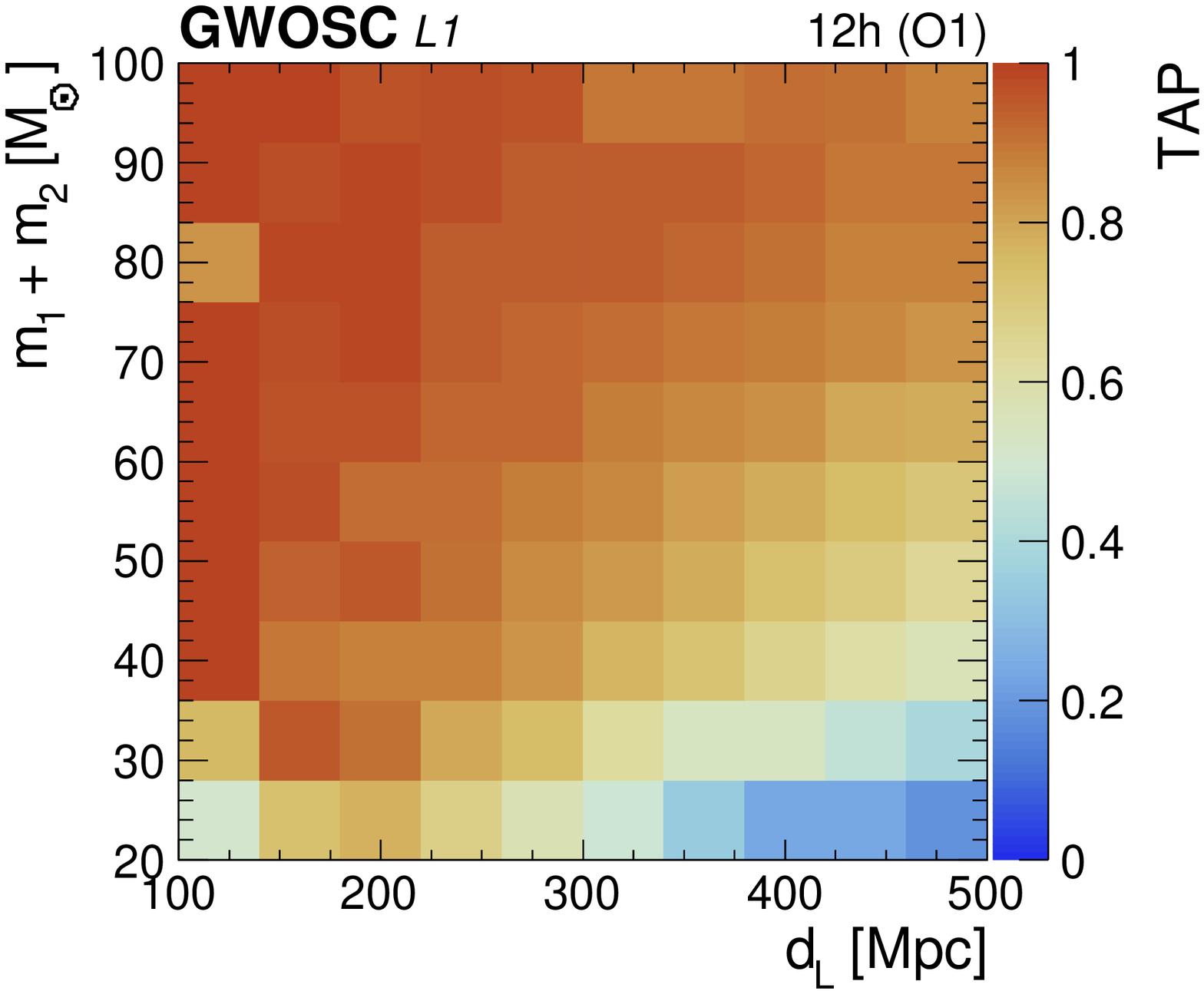}
\hspace{10pt}
\includegraphics[width=.45\linewidth]{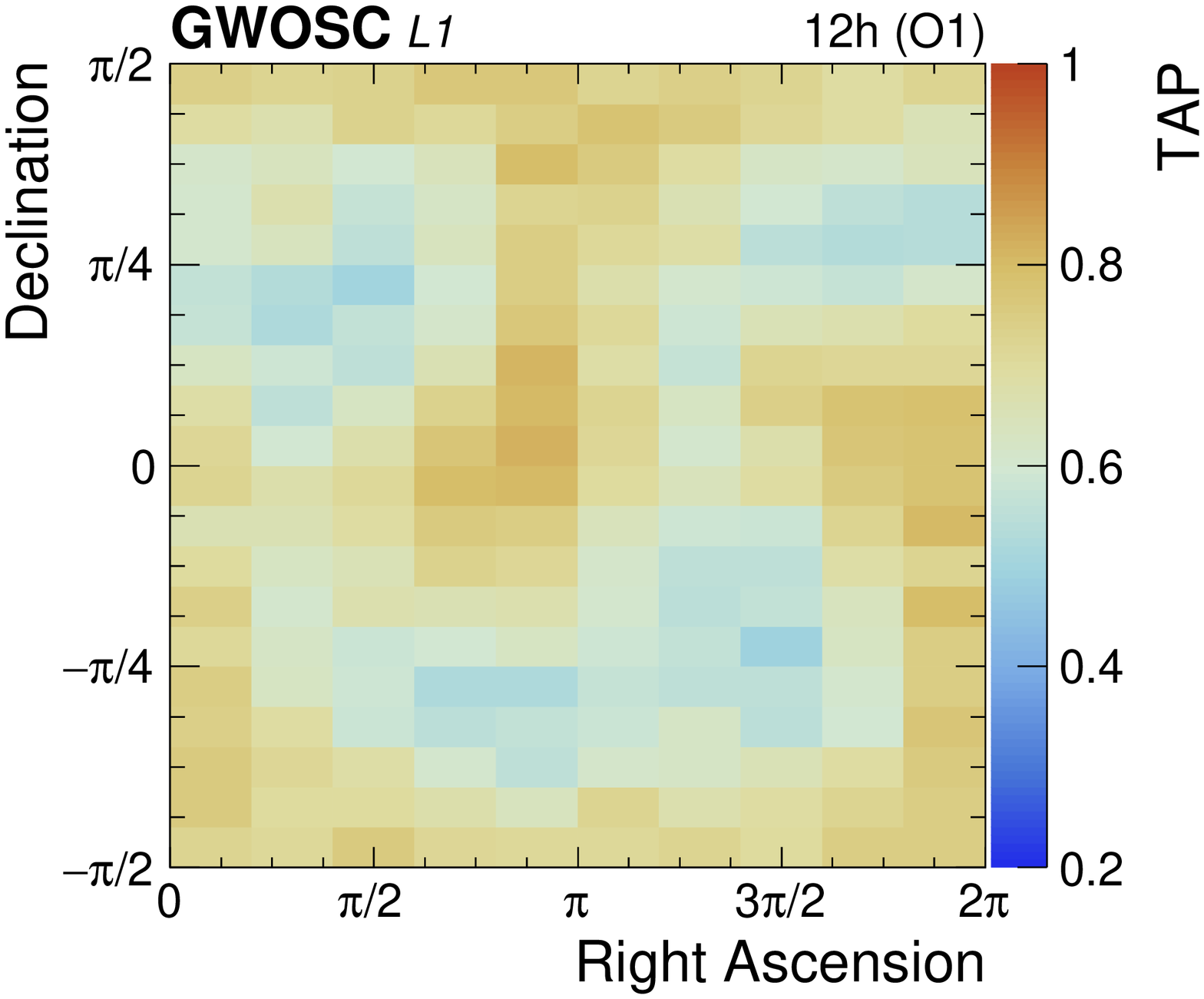}
\caption{The expected physics reach as a function of BBH observables for ML classifiers trained on inputs from H1+L1 (top), H1 (middle), and L1 (bottom). The TAP for a given bin, indicated by the color scale on the right of each plot, is for a FAP working point with a $3\sigma$ significance. Left: The TAP as a function of total BBH merger mass $\mtot$ vs. luminosity distance $\dL$. Sensitivity improves uniformly with multidetector inputs. Right: The TAP as a function of sky position. Multidetector inputs allow ML classifiers to capitalize on the independent information present in detectors of distinct angular response.}
\label{fig:phasespace}
\end{figure*}

Studying the improvement in detection sensitivity as a function of the BBH observables sheds light on what the H1+L1-trained ML classifier is learning over the 1-det ML classifiers. The expected physics reach for signal samples detected at the $3\sigma $ working point is plotted in Fig.~\ref{fig:phasespace} for H1 (middle) and L1 (bottom), in addition to those already introduced for H1+L1 (top). The TAP distribution in $\mtot$ versus $\dL$ (left) improves more or less uniformly with lower detector noise (H1 vs. L1) or an additional detector (H1+L1 vs. H1 or L1). The TAP distributions in sky position (right) illuminate the latter. Because of the differences in antenna orientation that the detectors present to a BBH source, the H1 (middle right) and L1 (bottom right) classifiers exhibit slightly different sensitivities in sky position. The L1 classifier, because of its higher associated detector noise in O1~\cite{ligoglitches2}, is also uniformly less sensitive than the H1 classifier in sky position, consistent with the earlier observation in $\mtot$ versus $\dL$. Thus, an essential component of what the H1+L1 classifier (top right) is learning is how to exploit independent angular information from disparate detectors. For any given region of sky, the stronger of the two detectors is leveraged. The RMS(H1,L1) ML classifier showed a similar improvement in angular sensitivity. An ideal array of GW detectors would thus be arranged to avoid blind spots in angular sensitivity over the sky. To avoid rejecting too many GW signals because of blind spots in any one detector, compatibility requirements between waveforms from different detectors could be modified to require only a majority agreement rather than complete agreement. Although we have not studied the response of the ML classifier relative to other BBH observables, investigating these may further illuminate what ML classifiers learn. In a later section (Section~\ref{sec:BBH:datavmc}), we examine whether the ML classifier is learning unrealistic artifacts of the simulation.

\subsection{Detector evolution}
\label{sec:BBH:datavtime}

\begin{figure*}
\centering
\includegraphics[width=.4\linewidth]{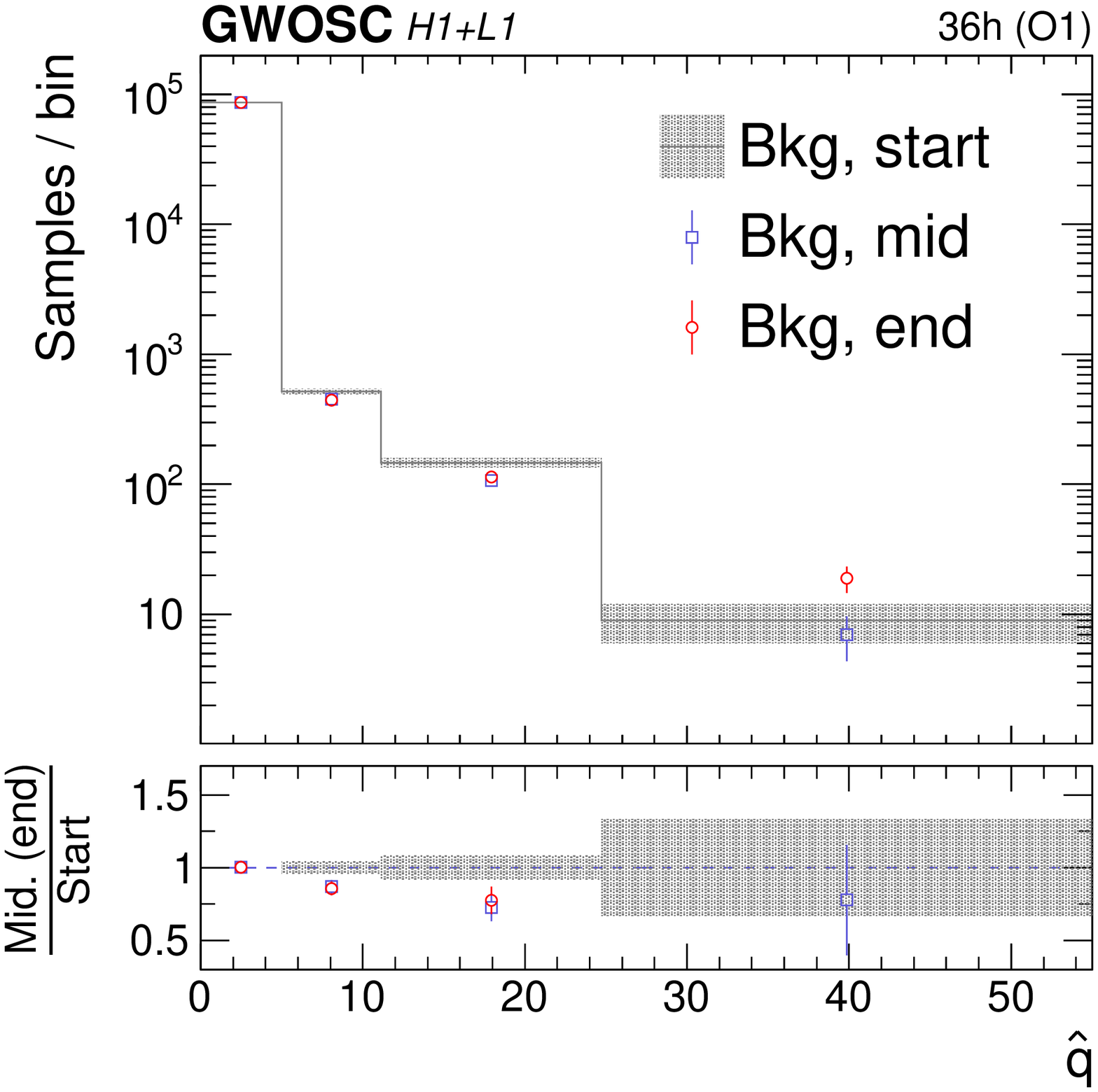}
\hspace{10pt}
\includegraphics[width=.4\linewidth]{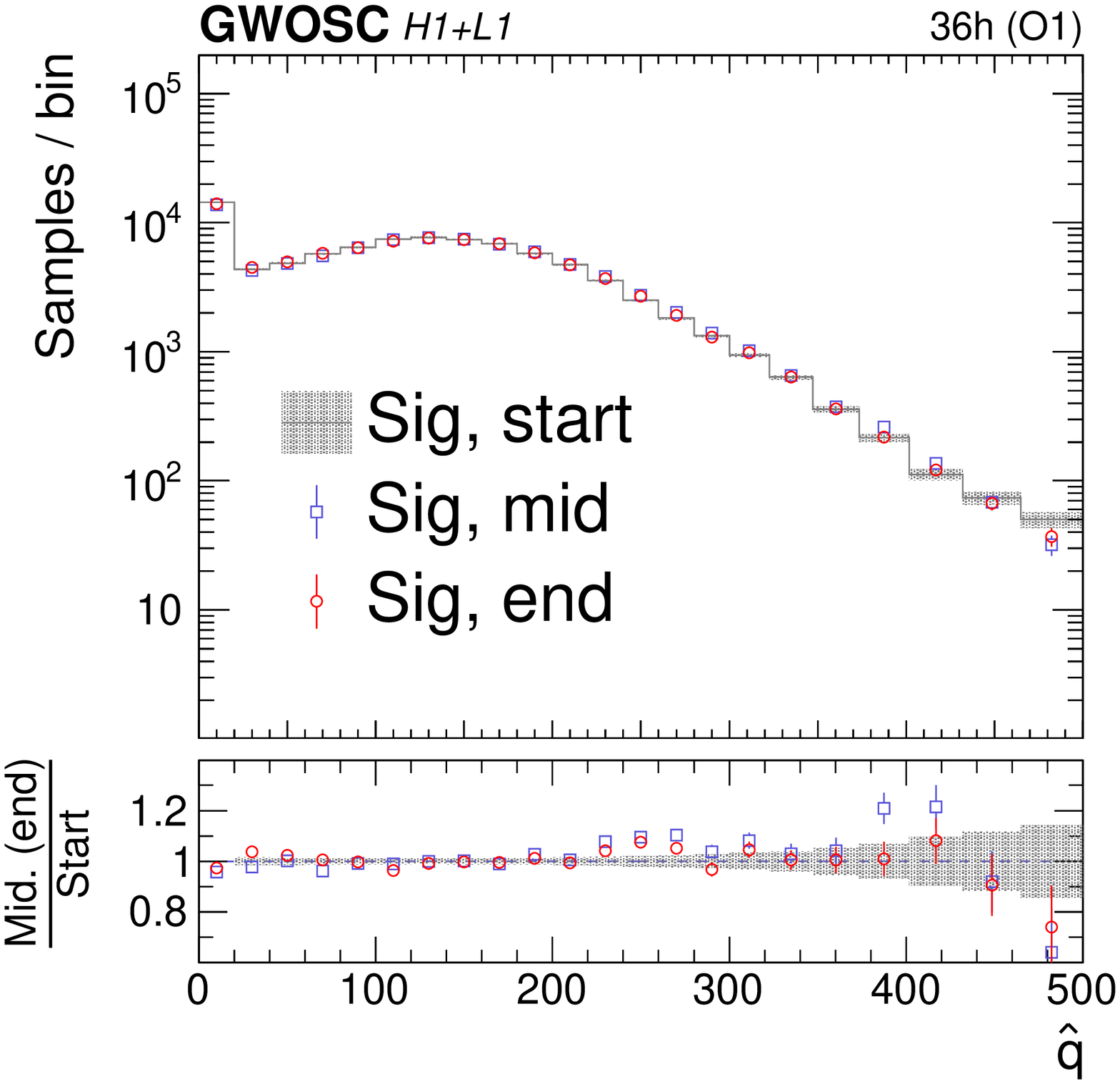}
\caption{Time evolution of the detection statistic $\qstat$ for background (Bkg, left) and signal (Sig, right) samples evaluated using the H1+L1-trained ML classifier. The upper panels plot the $\qstat$ distribution for time segments from the start (gray), middle (blue), and end (red) of O1 data-taking, with the filled band or vertical bars showing statistical uncertainties. The lower panel gives the ratio of the distributions for the middle to the start (blue squares) and the end to the start (red circles) of the O1 data-taking period, with vertical bars corresponding to statistical uncertainties in the numerator quantity. The gray band gives the similar uncertainty in the denominator quantity. Note the difference in x-axis scales between the left and right plots. The proportion of noise due to glitch-like events in the background distribution (left) decreases by 20–50\% causing percent-level changes in the signal distribution (right) for similar values of $\qstat$.}
\label{fig:nnoutvtime}
\end{figure*}

Next, we study the impact of evolving detector conditions on GW signal identification. We compare the $\qstat$ distribution from three time segments that correspond to the start, middle, and end of O1 data-taking. Since the ML classifier was trained using data from only the early part of O1, these studies double as checks for how well the ML classifier accommodates detector and noise conditions unseen in the training set. In Fig.~\ref{fig:nnoutvtime}, the $\qstat$ distribution for background (left) and signal samples (right) is shown overlaid for the three time segments. The proportion of noise due to glitch-like events in the right-side tail of the background distribution decreases by 20--50\% in the middle (blue) and end (red) of O1 data-taking relative to the start (gray). Note that we have not explicitly removed any glitches (other than those removed by the data quality flag requirement described in Section~\ref{sec:Data}). This change in noise distribution has a corresponding effect on signal sensitivity (right). The proportion of signal samples ranked with scores near the right-side tail of the background distribution increased by as much as 10\% while reducing the proportion of lower-scoring signal samples by up to a few percent. However, most of these changes occurred below the $3\sigma$ significance level. The TAP at the $3\sigma$ working point is thus stable to within 1\% between subsequent time segments (about two standard deviation of statistical uncertainty). Glitch mitigation, an important task for an MF-based pipeline, is therefore not as critical for an ML-based one.

\subsection{Simulation modeling}
\label{sec:BBH:datavmc}

Finally, we study the impact of simulation modeling on GW signal identification. We compare BBH signals discovered in O1 data (\textsc{GW150914}, \textsc{GW151012}, \textsc{GW151226}) from signals simulated with matching BBH parameters. These studies validate whether \deepsnr~is learning realistic GW features or is simply learning simulation artifacts. Such artifacts can arise from either generator mismodeling of the pure GW signal in \textsc{SEOBNRv4} or mismodeling of the detector response in \textsc{PyCBC}. Although the study we present is unable to unfold these individual effects, it does provide an upper bound for the effect of overall simulation mismodeling.

\begin{figure*}
\centering
\includegraphics[width=.325\linewidth]{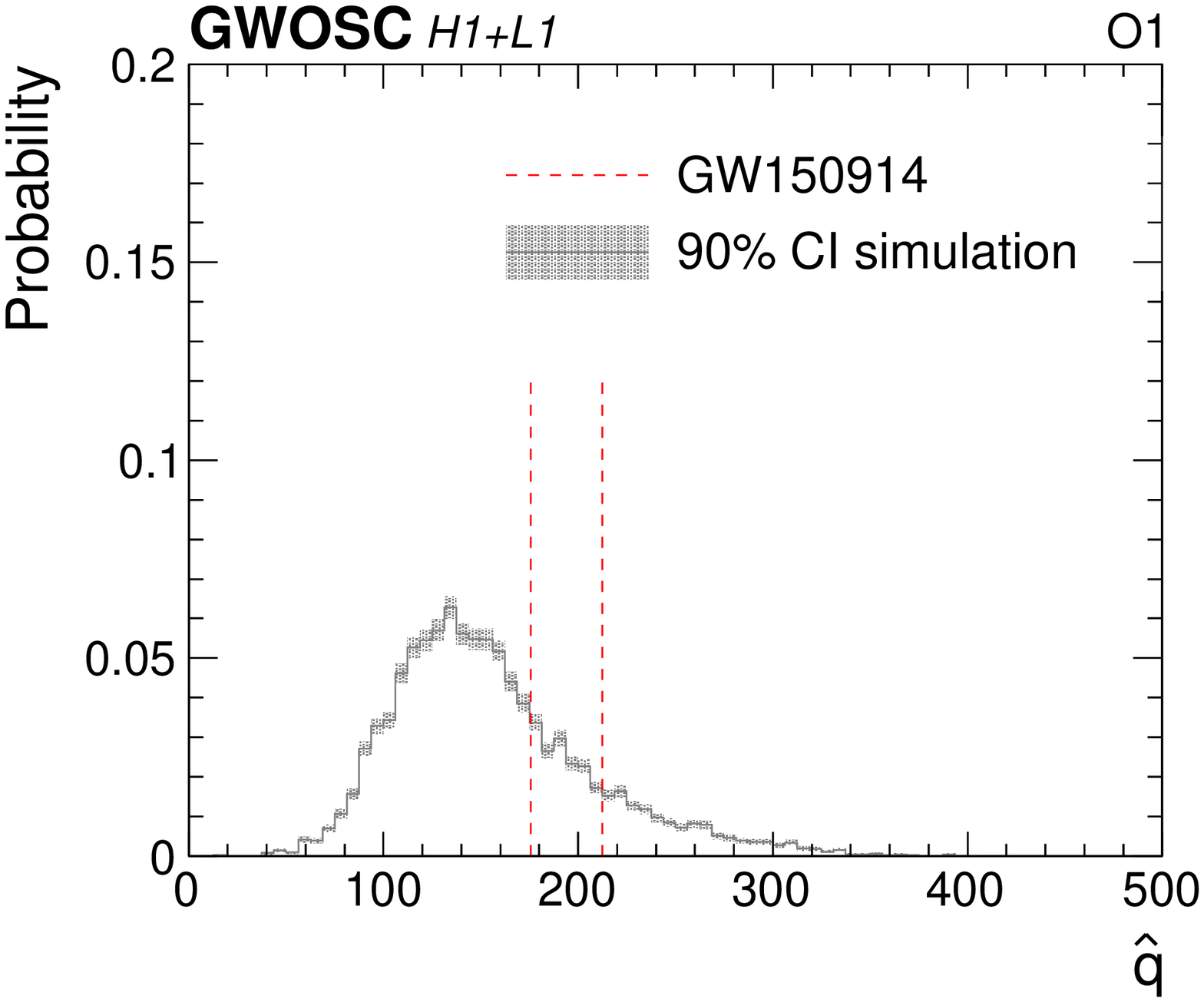}
\includegraphics[width=.325\linewidth]{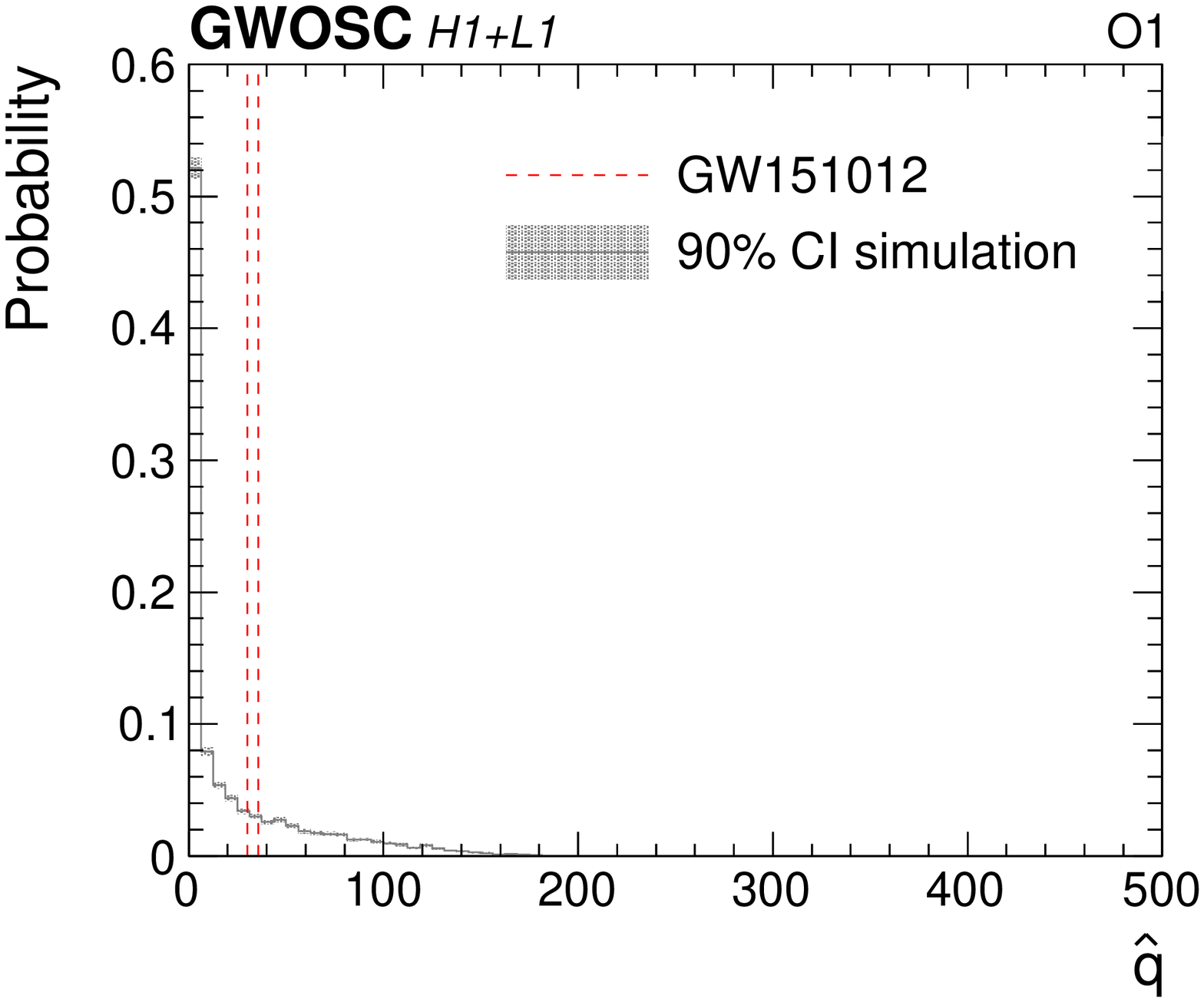}
\includegraphics[width=.325\linewidth]{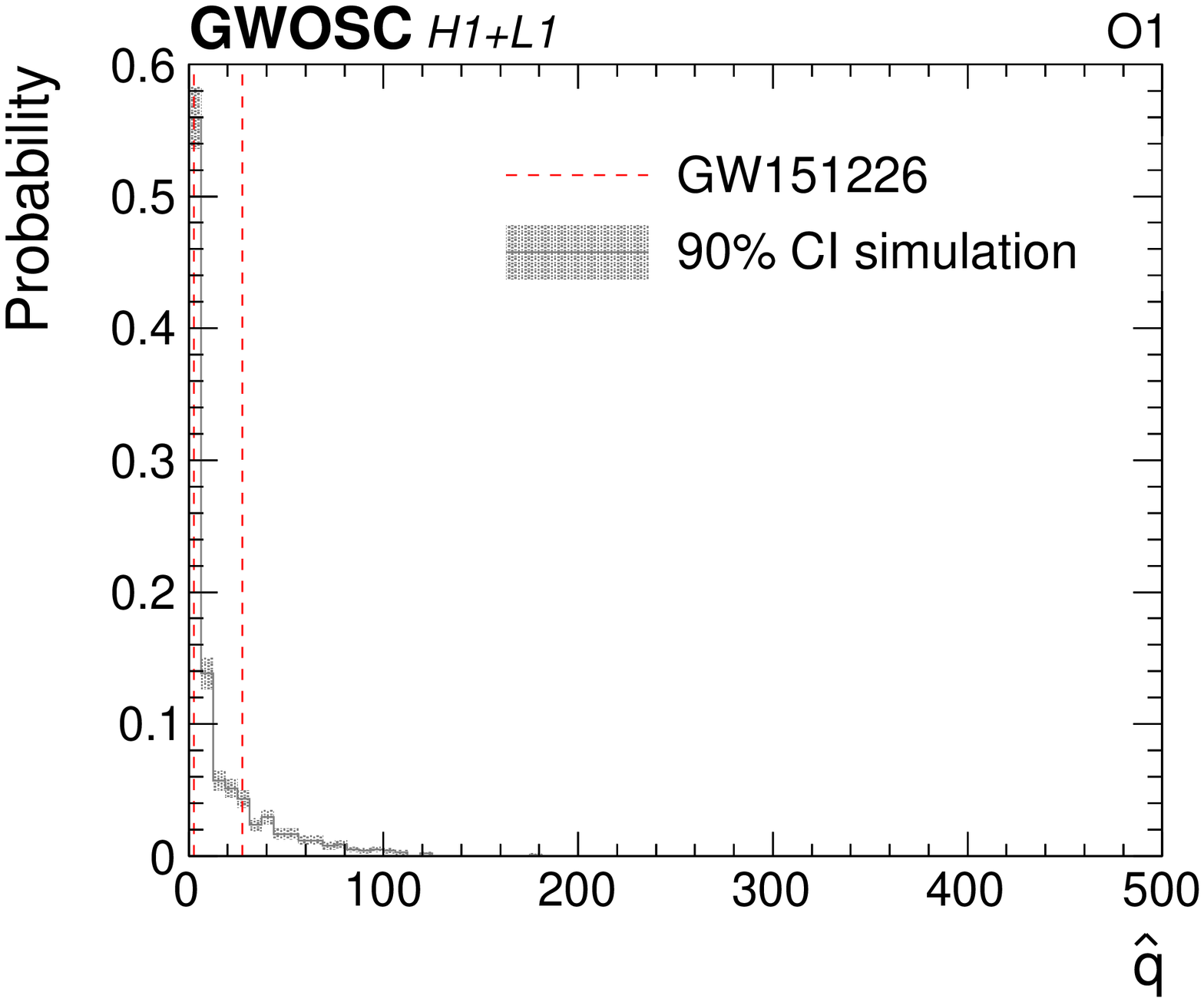}
\caption{Comparison of the detection statistic $\qstat$ in data (red vertical line) versus simulation (gray histogram) for the unclustered GW events \textsc{GW150914} (left), \textsc{GW151012} (middle), and \textsc{GW151226} (right). The simulated samples are generated using BBH posterior parameter distributions from the estimated 90\% CI of their data counterparts. The samples in data and simulation are evaluated using the H1+L1-trained ML classifier. The statistical uncertainties in the simulation are given by the gray band around the histogram. Note that the events are unclustered, and thus data events appear twice (c.f. Section~\ref{sec:Process}). The response in data is consistent with that in simulation at the 90\% CI.}
\label{fig:datavmc}
\end{figure*}

For each of the discovered events, we generate simulated BBH waveforms sampled from the 90\% CI posterior parameter distributions estimated for these events~\cite{gwtc1,gwosc}. Following the procedure described in Section~\ref{sec:Data}, the pure BBH waveforms are then randomly overlaid onto data-only segments in the 32~s around the identified GW event (excluding the actual GW event itself). In Fig.~\ref{fig:datavmc}, we plot the $\qstat$ distribution for the detected event versus its simulated counterpart. Note that the events are unclustered and thus appear twice as a consequence of how the time windows are created (described in Section~\ref{sec:Process}). For all three events, we find the detected signal to be well within the simulated distributions for the 90\% CI. Thus, inasmuch as can be concluded from this limited sample, we expect the \deepsnr~results from earlier sections based on simulated BBH signals to be relevant for the identification of BBH signals in real data. Note that a rigorous comparison of the detection significance for these events using \deepsnr~versus those reported by existing LIGO pipelines~\cite{gwtc1} is nontrivial and not possible with the current machinery we have developed, for the same reasons described in Sections~\ref{sec:Method:cands} and~\ref{sec:BBH:multidet}. We discuss the remaining challenges next.

\section{Discussion}
\label{sec:Discussion}

Our results with \deepsnr~demonstrate, for the first time, the feasibility of using ML to enable the potential discovery of previously unobserved GW signals in offline detection pipelines in experiments at LIGO. It must be cautioned, however, that \deepsnr~is only a foundation for a fully MF-free offline GW-detection pipeline. Additional machinery is required to perform the ultimate goal of significance quantification as it is done in current LIGO GW searches. A helpful illustration of the \textsc{PyCBC} offline detection pipeline can be found in Ref.~\cite{magicbullet}. In particular, the ML classifier of \deepsnr~has no mechanism to ascertain the compatibility of candidate waveforms from different detectors---the so-called coincidence requirement~\cite{gwtc1,pycbc}. Even though \deepsnr~exploits information from both detectors for signal classification, it does not explicitly require compatibility. Waveform compatibility is an important and robust layer of protection against false alarms provided by the MF algorithm. The matched templates from the MF scan serve as handles for comparing both the arrival times and BBH parameters of candidate waveforms between detectors. Moreover, the coincidence rate of matching templates between detectors is used as a baseline for measuring the real-world significance of a GW signal candidate. An ML-based counterpart for this step can be achieved using a parameter regressor similar to those developed in Refs.~\cite{paramestinterval,paramestnormflow1,paramestnormflow2}. Signal candidates identified by \deepsnr~would then be run through the ML parameter regressor to determine the arrival times, best-fit parameters, and associated confidence intervals of the waveforms from different detectors. An equivalent coincidence requirement in the ML-based scheme can then be achieved by requesting consistent arrival times and overlapping confidence intervals for the regressed parameters. A version of this workflow was described in Ref.~\cite{huertabbh-id-param}. After this step, clusters of identified signal candidates passing and failing the compatibility requirements can then be used to build a distribution for estimating the final background and thus the detection significance. We expect that existing LIGO analysis techniques developed for other SNR-like detection will be usable without major modification. The investigation of such a fully ML-based offline detection pipeline is planned for future work.

A comparison between a completely ML-based offline detection pipeline with a real-world MF benchmark would be needed to bear out the ultimate value of an ML-based approach. An important potential benchmark application in this regard is the detection of BNS mergers. Because of the complex tidal interactions involved in BNS mergers, their dynamics are difficult to probe fully within the computational constraints of MF scans.
Moreover, ML-based approaches are likely to be more robust than MF-based ones at extrapolating to BNS waveforms that cannot be modeled at all.
Early ML work in this direction using simplified models has shown promise~\cite{harvardbns-id-param,harvardbns-id}. Follow-up studies using more realistic simulations and detector responses similar to those we have presented for BBH mergers will be important tests.

More generally, the computational advantage of ML- over MF-based detection makes it more practical to probe different GW signal types with increased sensitivity. For each GW signal type, the time complexity of running an MF scan scales like $\mathcal{O}(N_{\mathrm{data}} \, N_{\mathrm{params}}\, N_{\mathrm{steps}})$, where $N_{\mathrm{data}}$ is the number of GW candidates to be evaluated (data set size), and $N_{\mathrm{params}}$ and $N_{\mathrm{steps}}$ are, respectively, the number of signal parameters (fidelity) and steps in each parameter (resolution) used to build the MF template bank. This polynomial time complexity makes it computationally expensive to probe a wide variety of GW signal categories with high parameter fidelity and resolution, especially with large data sets. By contrast, an ML-based classifier like \deepsnr~can be expanded to identify multiple GW signal classes simultaneously so that each candidate waveform in the data set is evaluated with constant time complexity or $\mathcal{O}(N_{\mathrm{data}})$ over the full data set to be evaluated. Although the training set will be correspondingly larger, the cost of training is only incurred once.  This makes it more practical to train on a set of waveforms with better parameter fidelity and resolution. Moreover, when training with time-domain inputs as with \deepsnr, the cost of running a Fourier Transform to obtain frequency-domain information is entirely avoided. Since multiclass ML classifiers are likely to learn more diverse and nuanced waveform features, we expect these classifiers to effectively extrapolate to unseen waveforms that cannot be easily modeled and simulated, even for known signal types. This raises the possibility of building an unsupervised ML classifier that can detect GWs of unknown signal type by using the features learned from the pretrained supervised classifier. Caution, however, must again be exercised to ensure that any such ML classifiers, supervised or unsupervised, are properly validated and understood.

For current and future GW detectors like KAGRA~\cite{KAGRA}, LISA~\cite{LISA}, DECIGO~\cite{DECIGO}, TianQin~\cite{TianQin}, Cosmic Explorer~\cite{CE}, Einstein Telescope~\cite{ET}, and the Big Bang Observer~\cite{BBO}, the potential of ML-based offline detection hinted at by \deepsnr~makes it important for GW collaborations to invest in the serious study of fully ML-based offline detection pipelines. In the near term, it is not necessary to develop completely new offline detection pipelines. Instead, existing pipelines like \textsc{PyCBC}~\cite{pycbc} and \textsc{GstLAL}~\cite{gstlal} can simply be retooled to use an ML-based foundation instead of an MF-based one. This would provide an important first step in validating ML-based offline detection pipelines against established MF-based ones. In the longer term, collaborations can design and optimize the full offline detection pipeline to best exploit the strengths of the underlying ML foundation.

Although we have proposed \deepsnr~for offline detection, we expect it to be useful for online detection as well. Even if using a simple background model as we have, ranking potential signal candidates based on their $\qstat$ (corresponding to a given FAP) would provide a more interpretable quantification of the significance of an online signal candidate over a probability-based score. The GW ML literature targeting online detection~\cite{huertadeepfilter,harvardbns-id,mitrealtime,huertahpc}, all of which utilize a probability score formulation, thus stands to benefit from the SNR-based formulation we describe.

Lastly, although we defined the detection statistic $\qstat$ in the context of GW detection, we expect that such an interpretation of ML classification scores would be of general value in any ML classification task where one is interested in the statistical analysis of a distribution of rare, signal-like outliers against some background process. Such applications might include the experimental identification of other, rare astronomical events, like neutrino interactions.

\section{Summary}
\label{sec:Summary}

The \deepsnr~detection pipeline is introduced, which uses a novel method for generating an SNR ranking statistic from deep learning classifiers, providing for the first time a foundation for powerful and computationally efficient deep learning algorithms to be leveraged in the potential discovery of weaker and never-before-observed GW signals. \deepsnr~is based on a \textsc{ResNet} convolutional neural network classifier simultaneously trained on time-domain waveforms from both the Hanford and Livingston LIGO detectors. Open LIGO data collected during the first observation run and simulated GW signals with realistic detector response are used.

A novel function is defined over the raw output of the ML classifier to construct the SNR detection statistic for ranking signal candidates against background samples. The statistic emphasizes the distribution of signal-like outliers against the distribution of noise sources to facilitate statistical analysis with existing techniques used in matched-filter-based offline detection pipelines. Using this detection statistic, a strategy for identifying potential signal candidates based on the false alarm probability relative to a simple background model is proposed. To validate the \deepsnr~pipeline, its use is demonstrated in the identification of BBH mergers in open LIGO data. The expected sensitivity to BBH sources is quantified using the true alarm probability at various false alarm probability working points. Using high-fidelity simulations of the LIGO detector responses, the expected physics reach---with respect to BBH observables including merger mass, luminosity distance, and sky position---is presented for the first time.

The impact of the number of LIGO detectors, detector noise evolution, and simulation modeling on \deepsnr~is also quantified in an ML first. A benchmark against a simplified matched-filter algorithm without any glitch mitigation is performed for context. Sensitivity gains of several orders of magnitude are achieved at a fraction of the computational cost, highlighting the intrinsic advantage of \deepsnr~at glitch mitigation and computational efficiency. As seen when given a varied number of detector inputs, \deepsnr~has the key advantage of being able to exploit independent information from detectors with distinct angular responses, a feature not seen in matched-filter algorithms.

Signal identification with \deepsnr~is found to be robust under a detector noise evolution of order months. The response of \deepsnr~to simulated versus observed GW signals is consistent at the 90\% confidence interval. These studies pave the way toward an offline GW-detection pipeline that exploits the strengths of deep learning to extend the reach of GW detectors, current and future, and enable the potential discovery of challenging, never-before-observed GW signals.

\begin{acknowledgments}
We would like to thank Adam Lewis, Edward Rietman, Philip Chang, and Jerry Tessendorf for helpful comments and valuable discussions at different stages of this study. The work of MA and MP is supported in part by the Department of Energy, Office of Science, through DOE award DE-SC0010118 and by the National Science Foundation through the NSF AI Planning Institute: Physics of the Future, NSF PHY-2020295.

This research has made use of data or software obtained from the Gravitational Wave Open Science Center (\texttt{gw-openscience.org}), a service of LIGO Laboratory, the LIGO Scientific Collaboration, the Virgo Collaboration, and KAGRA. LIGO Laboratory and Advanced LIGO are funded by the United States National Science Foundation (NSF) as well as the Science and Technology Facilities Council (STFC) of the United Kingdom, the Max-Planck-Society (MPS), and the State of Niedersachsen/Germany for support of the construction of Advanced LIGO and construction and operation of the GEO600 detector. Additional support for Advanced LIGO was provided by the Australian Research Council. Virgo is funded, through the European Gravitational Observatory (EGO), by the French Centre National de Recherche Scientifique (CNRS), the Italian Istituto Nazionale di Fisica Nucleare (INFN) and the Dutch Nikhef, with contributions by institutions from Belgium, Germany, Greece, Hungary, Ireland, Japan, Monaco, Poland, Portugal, Spain. The construction and operation of KAGRA are funded by Ministry of Education, Culture, Sports, Science and Technology (MEXT), and Japan Society for the Promotion of Science (JSPS), National Research Foundation (NRF) and Ministry of Science and ICT (MSIT) in Korea, Academia Sinica (AS) and the Ministry of Science and Technology (MoST) in Taiwan.
\end{acknowledgments}

\appendix
\section{Supplementary studies}

\renewcommand{\thefigure}{A.\arabic{figure}}
\renewcommand{\thetable}{A.\arabic{table}}
\subsection{Time versus frequency domain}
\label{app:tvf}

A principal question in the study of GW data---and waveform data, in general---is whether to analyze the time-domain strains or the Fourier-transformed, frequency-domain spectrograms (i.e., frequency vs time graphs)~\cite{ligomlrev}. Time-domain strains prove simpler when processing and training ML classifiers but can contain frequencies dominated by detector noise, potentially making the extraction of signal information difficult. On the other hand, creating and working with spectrograms is computationally expensive but has the potential advantage of presenting decomposed frequency information.

In Table~\ref{tab:tvf}, we compare the signal sensitivity of ML classifiers trained on time- or frequency-domain inputs. The frequency-domain spectrograms are constructed from the raw time-domain strains using a short-time Fourier Transform (FFT)~\cite{qtransform,gwpysw} with whitening. The FFT parameters are chosen to output spectrograms with a frequency and time resolution of $\Delta f\times\Delta t = (1.6~\mathrm{Hz},~ 2~\mathrm{kHz^{-1}})$ or array dimensions $(205,178)$ for a time window of 1~s and frequency range of 32 to 320 Hz. Different training set sizes are compared to account for potential differences in learning curves. The results indicate that ML classifiers trained on time-domain inputs are significantly more sensitive than those trained frequency-domain inputs. We note, however, that the use of spectrograms introduces several additional processing and ML model parameters, in addition to computational cost. Although we have not performed an exhaustive optimization of these additional parameters, we find no evidence that using frequency-domain inputs carries any sensitivity advantage over the much simpler time-domain inputs. As summarized below, we find the results of Table~\ref{tab:tvf} to be additionally robust with respect to data processing. Thus, without loss in signal sensitivity, the studies performed in the main body of this paper use only time-domain inputs.

\begin{table}[!htbp]
\caption{Signal sensitivity of ML classifiers trained on time-domain waveforms (left, $t$ domain) versus frequency-versus-time spectrograms (right, $f$ domain) using inputs from H1 (top), L1 (middle), or both H1+L1 (bottom). For each input type, the results are broken down by training set size, of number $0.5\ntrain$ / $\ntrain$ / 2$\ntrain$, where $\ntrain=95$k is the nominal training set size. The detection sensitivity is expressed in terms of the TAP at a FAP working point with $3\sigma$ significance (TAP@FAP:3$\sigma$). The values are measured using the same validation set of size $\nval=45$k, corresponding to statistical uncertainties of $\pm0.005$. Uncertainties due to the stochastic nature of ML training are a few percent. The results favor time-domain inputs, a conclusion we find to be robust versus training set size for H1+L1 inputs.}
\label{tab:tvf}
\begin{ruledtabular}
\begin{tabular}{ccc}
 &\multicolumn{2}{c}{TAP@FAP:$3\sigma$}\\
 &\multicolumn{2}{c}{0.5$\ntrain$ / $\ntrain$ / 2$\ntrain$}\\
 & $t$ domain & $f$ domain \\
 \hline
 H1 only  & 0.87 / 0.81 / 0.89 & 0.77 / 0.78 / 0.79 \\
 L1 only  & 0.83 / 0.78 / 0.84 & 0.71 / 0.71 / 0.72 \\
 H1+L1    & 0.89 / 0.91 / 0.91 & 0.84 / 0.85 / 0.85 \\
\end{tabular}
\end{ruledtabular}
\end{table}

\paragraph*{Time-domain processing.}
\label{app:trobust}

To assess the robustness of the time-domain results in Table~\ref{tab:tvf}, we study the effect of varying the band-pass filter range and the time window used to convert the continuous strain data to array segments for training (described in Section~\ref{sec:Data}). The optimal band-pass range should include only frequencies where the power of the sought GW signal is above that of the detector noise. However, since the power spectrum of GW signals varies with merger mass, a good measure of the robustness of time-domain-based results is the change in signal sensitivity if the training is performed on waveforms processed with looser band-pass filter ranges. Compared with the time-domain ML classifier trained on H1+L1 waveforms with nominal frequencies between 32 and 320 Hz, those trained on waveforms with band pass ranges of 32--512 Hz and 16--2048 Hz have signal sensitivity a few percent and 15\% worse, respectively, at the $3\sigma$ significance level. Our time-domain results are thus likely to be no more than a few percent from optimal for the chosen BBH parameter range. Optimal results may potentially be achieved by learning which frequencies should be removed---for instance, by using an autoencoder~\cite{gwdenoise}. We note that the ML classifier trained on waveforms with no band-pass filter and no whitening applied failed to converge.

The optimal time window for time-domain-based ML analysis is one that sufficiently accommodates the duration over which the GW signal strains are above those of typical detector noise. Although BBH merger signals may last several seconds for the masses we study, their strains are above those of detector noise for only fractions of a second. We confirm this by varying the time window from 1~s to 0.5~s and 2~s. We find the time-domain ML classifiers trained on H1+L1 inputs with 1~s time windows to have signal sensitivity comparable to those trained with 0.5~s and 2~s time windows, within uncertainties. Thus, the choice of time window is approximately optimal. For longer time windows, we note that the probability for rarer noise artifacts to contaminate the input sample increases correspondingly.

\paragraph*{Frequency-domain processing.}
\label{app:frobust}

To assess the robustness of the frequency-domain results in Table~\ref{tab:tvf}, we study the effect of adjusting the FFT processing to give twice the time and frequency resolution of the resulting spectrograms. That is, we increase the spectrogram resolution in frequency and in time from the nominal $\Delta f\times\Delta t = (1.6~\mathrm{Hz},~2~\mathrm{kHz^{-1}})$ to $(0.8~\mathrm{Hz},~4~\mathrm{kHz^{-1}})$, corresponding to spectrogram dimensions $(205,178)$ to $(410,355)$, respectively. The sensitivity of the ML classifier trained on H1+L1 inputs at the increased spectrogram resolution is consistent with that of the nominally trained one within uncertainties. We therefore expect the frequency-domain results of the H1+L1 training in Table~\ref{tab:tvf} to be close to optimal with respect to processing parameters. We do not, however, consider the effect of varying the ML model parameters.

\subsection{ML architecture optimization}
\label{app:arch}

We experimented with different 1-dimensional implementations of the original \textsc{ResNet} CNN architecture~\cite{resnet}. All implementations used \textsc{ReLU} activation functions and no batch normalization. Batch normalization tends to be destructive because GW detection is sensitive to the overall strain scale. Since the input array length (4096) is quite large, an efficient scheme for reducing and extracting information without overtraining is critical. These can be achieved either by aggressively downsampling the array before passing it to the classifier (pre-downsampled), or by making liberal use of max-pooling (pooling heavy) or using larger striding (stride heavy) in the classifier model itself. The pooling- and stride-heavy strategies have comparable sensitivity, within uncertainties, while the pre-downsampled strategy is worse by several percent. This is to be expected because downsampling within the ML model itself allows information to be extracted even as the inputs are reduced. A stride-heavy strategy (convolutional kernels of size $7\!\!\times\!\!7$ and stride 3) has the advantage of having less model parameters than the pooling-heavy one because fewer layers are needed to achieve the same input reduction. Since the features in the GW tend to span several array segments, as long as a sufficiently large window is used to perform the convolutions, the risk that an important feature is missed by using a larger stride is negligible. Thus, for all the studies presented in this paper, the stride-heavy \textsc{ResNet} model is used exclusively. A \textsc{ResNet} implementation with 10 (H1+L1 inputs) or 20 (H1 or L1 inputs) convolutional layers is found to be most performant. Implementations written in \textsc{PyTorch} of the various \textsc{ResNet} models described above can be found in Ref.~\cite{mygithub}.

\end{document}